 \documentclass[smallabstract,smallcaptions]{dccpaper}

\usepackage{epsfig}
\usepackage{citesort}
\usepackage{amsmath}
\usepackage{amssymb}
\usepackage{color}
\usepackage{url}

\newlength{\figurewidth}
\newlength{\smallfigurewidth}

\usepackage{caption}
\usepackage{subcaption}
\usepackage{soul}
\usepackage{float}
\usepackage{cancel}
\usepackage{tikz}

\usepackage{relsize}

\usepackage[affil-it]{authblk}

\usetikzlibrary{backgrounds,calc,chains,fit,matrix,positioning,shadows,shapes.misc,circuits.ee}
\tikzset{input/.style={}}
\tikzset{output/.style={}}
\tikzset{operator/.style={circle, draw, minimum size=2.5ex, inner sep=0pt}}
\tikzset{filter/.style={rectangle, draw, fill=white, minimum size=3.5ex, inner xsep=1.5ex}}
\tikzset{filter2/.style={rectangle, draw, fill=white, minimum size=3.5ex, inner xsep=3.5ex, inner ysep=3.5ex}}
\tikzset{other/.style={rounded rectangle, draw, fill=white, minimum size=3.5ex, inner xsep=1ex}}
\tikzset{branch/.style={circle, draw, fill=black, thick, minimum size=.5ex, inner sep=0pt}}
\tikzset{rv/.style={circle, draw, thick, fill=white, minimum size=2.75ex, inner sep=0pt}}
\tikzset{ob/.style={circle, draw, thick, fill=lightgray, minimum size=2.75ex, inner sep=0pt}}
\tikzset{pa/.style={circle, draw, thick, fill=black, minimum size=1ex, inner sep=0pt}}
\tikzset{/tikz/thin/.style={line width=.6pt}}
\tikzset{/tikz/thick/.style={line width=1pt}}
\tikzset{>=direction ee}

\usepackage{hyperref}

\usepackage{cite}

\usepackage{xr}
\usepackage{amsmath}
\usepackage[cmintegrals]{newtxmath}

\usepackage{amsfonts}

\usepackage{caption}
\usepackage{subcaption}
\usepackage{xcolor}
\begin{document}

\title
{\large
\textbf{Neural Distributed Image Compression using Common Information}
}


\author{%
Nitish Mital$^{1, {\ast}}$, Ezgi {\"O}zyılkan$^{1, 2, {\dag}}$, Ali Garjani$^{1, 2, {\ddag}}$, and Deniz G{\"u}nd{\"u}z$^{\ast}$\\
{\small\begin{minipage}{\linewidth}\begin{center}
\begin{tabular}{c}
$^{\ast}$Dept.~of Electrical and Electronics Engineering, Imperial College London \\
$^{\dag}$Dept.~of Electrical and Computer Engineering, New York University \\
 $^{\ddag}$Section of Mathematics, EPFL  \\
\{n.mital, d.gunduz\}@imperial.ac.uk, eo2135@nyu.edu, ali.garjani@epfl.ch
\end{tabular}
\end{center}\end{minipage}}
}

\footnotetext[1]{Contributed equally to this work.}
\footnotetext[2]{At the time of this work, E. {\"O}zyılkan and A. Garjani were with the Dept.~of Electrical and Electronics Engineering, Imperial College London, and  Dept.~of Computer Engineering, Sharif University of Technology, respectively.}

\maketitle

\begin{abstract}
We present a novel deep neural network (DNN) architecture for compressing an image when a correlated image is available as side information only at the decoder. This problem is known as distributed source coding (DSC) in information theory. In particular, we consider a pair of stereo images, which generally have high correlation with each other due to overlapping fields of view, and assume that one image of the pair is to be compressed and transmitted, while the other image is available only at the decoder. In the proposed architecture, the encoder maps the input image to a latent space, quantizes the latent representation, and compresses it using entropy coding. The decoder is trained to extract the common information between the input image and the correlated image, using only the latter. The received latent representation and the locally generated common information are passed through a decoder network to obtain an enhanced reconstruction of the input image. The common information provides a succinct representation of the relevant information at the receiver. We train and demonstrate the effectiveness of the proposed approach on the KITTI and Cityscape datasets of stereo image pairs. Our results show that the proposed architecture is capable of exploiting the decoder-only side information, and outperforms previous work on stereo image compression with decoder side information.
\end{abstract}

\section{Introduction}

Data compression is a fundamental and well-studied problem in engineering, and is commonly formulated with the goal of designing codes with minimal average code length for a given data ensemble. Shannon showed that the entropy is a fundamental bound in lossless data compression when multiple independent samples of the information source can be compressed jointly while allowing arbitrarily small probability of error. The design of entropy codes approaching the Shannon limit relies on modeling the probability distribution of the data ensemble. Continuous-valued data (such as vectors of image pixel intensities) must also be quantized to a finite set of discrete values, which introduces error. In this context, known as the lossy compression problem, one must trade off two competing costs: the entropy of the discretized representation (rate) and the error arising from the quantization (distortion). In the case of lossy compression, the fundamental performance bound is characterized by the information theoretic rate-distortion curve. In practice, lossy compression is a challenging problem, and codes are designed separately for specific information sources, e.g., image, audio, video. In lossy image compression, practical codes typically follow a two-step approach, where a linear transform is followed by quantization and lossless compression using entropy coding (e.g., JPEG, JPEG 2000). Recently, deep neural network (DNN) aided data-driven image compression algorithms have received significant research interest \cite{balle2017, balle2018,8100060,theis2017lossy,NEURIPS2018_53edebc5,lee2018contextadaptive,Patel_2021_WACV}, and achieved impressive performance results, outperforming classical methods, such as JPEG 2000 and BPG \cite{952804,bpg}. 

\begin{figure}
    \centering
 \includegraphics[scale=0.4]{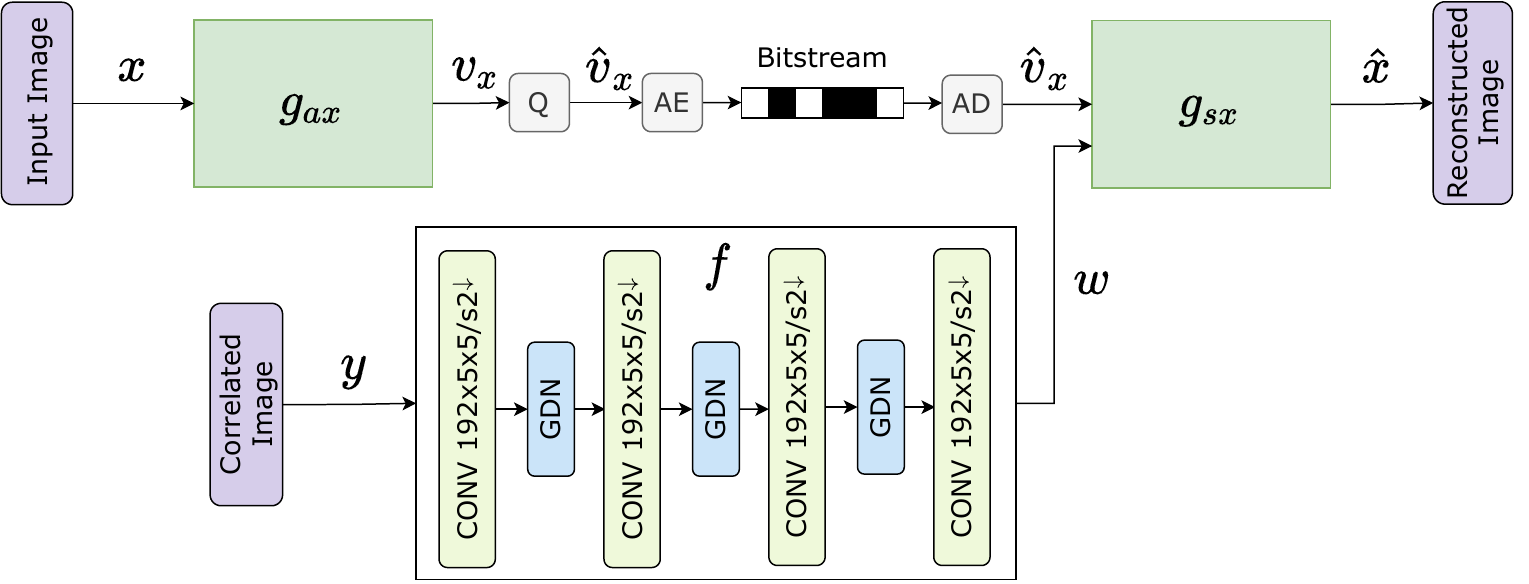}
    \caption{The proposed network architecture for distributed source coding. Block Q corresponds to a uniform quantizer, while blocks AE and AD correspond to arithmetic encoder and arithmetic decoder, respectively.} 
    \label{fig:dsc_architecture2}
\end{figure}

In this work, we are interested in DNN-aided \textit{distributed image compression}, where side information in the form of a correlated image is available only at the decoder. This scenario, illustrated in Fig.~\ref{fig:dsc_architecture2}, occurs, for example, in the case of a pair of stereo cameras, where two cameras capture images of a scene from different angles at the same moment. In this case, the two images are highly correlated due to overlapping fields of view. The two cameras do not communicate with each other, and therefore, cannot apply joint encoding of the images as in \cite{Liu_2019_ICCV}. Assume that the left camera delivers its image (in a lossless fashion) to the destination, e.g., a central storage or processing unit. The right camera, instead of employing a standard image compression algorithm, should be able to benefit from the presence of a highly correlated image from the left camera, even though it does not have access to this image. 
The benefit of decoder-only side information in compression goes back to Slepian and Wolf's seminal work on distributed lossless compression \cite{Slepian:IT:73}. They showed that a compression rate equal to the entropy of the source conditioned on the side information is necessary and sufficient for lossless compression, which is, interestingly, the same rate when the side information is available to both the encoder and the decoder. This was later extended by Wyner and Ziv to lossy compression with decoder side information in \cite{Wyner:IT:76}, which provides unbounded gains in rate; however, there is a non-zero rate loss in general compared to having the side information both at the encoder and the decoder.

In this paper, we propose a novel neural architecture to perform lossy compression of an image assuming that its stereo pair is available at the decoder as side information, similarly to \cite{DSIN}. A related work that also studies neural distributed source coding, \cite{whang2021neural}, considers a different setup where the bottom half of the image of a face is used as side information for reconstructing the upper half of the face. The novelty of the proposed method lies in employing the concept of common information to first learn the common features between the two correlated images. The common information generated by the decoder is then used along with the quantized latent representation of the original image conveyed by the encoder to reconstruct the original image. We show that our proposed method achieves a significantly better rate-distortion trade-off at low bit rates than the state-of-the-art single image compression algorithms, as well as previous work on deep stereo image compression with decoder side information \cite{DSIN}.

\textbf{Wyner-Ziv compression: }Let $X$ and $Y$ denote the source and side information variables, respectively, with joint distribution $p(x,y)$. The information theoretic fundamental limit in lossy compression is characterized by the \textit{rate-distortion function}, given by $R^{WZ}_{X|Y}(d) = \inf I(X;V\mid Y)$,
where $R^*(d)$ is the rate of compression (in bits per source sample) to achieve a given target average distortion $d$, between the input sequence $\mathbf{x} \in \mathbb{R}^n$ and its reconstruction $\mathbf{\hat{x}} \in \mathbb{R}^n$. The infimum is with respect to all auxiliary random variables $V$ and reconstruction functions $f: \mathcal{Y} \times \mathcal{V} \rightarrow \hat{\mathcal{X}}$ that satisfy: i) $V$ and $Y$ are conditionally independent given $X$, that is, $V - X - Y$ form a Markov chain; ii) $\mathbb{E}\left[ D(X,f(V,Y)) \right] \leq d$.

\begin{figure}[]
\centering
\begin{minipage}[b]{0.41\linewidth} 
\centering
\includegraphics[scale=0.55]{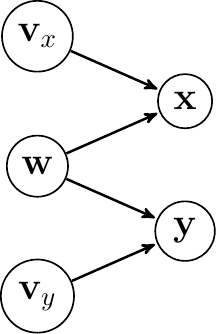} 
\caption{Graphical model.}
\label{fig:prob_model}
\end{minipage}
\quad
\begin{minipage}[b]{0.55\linewidth}
\includegraphics[scale=0.4]{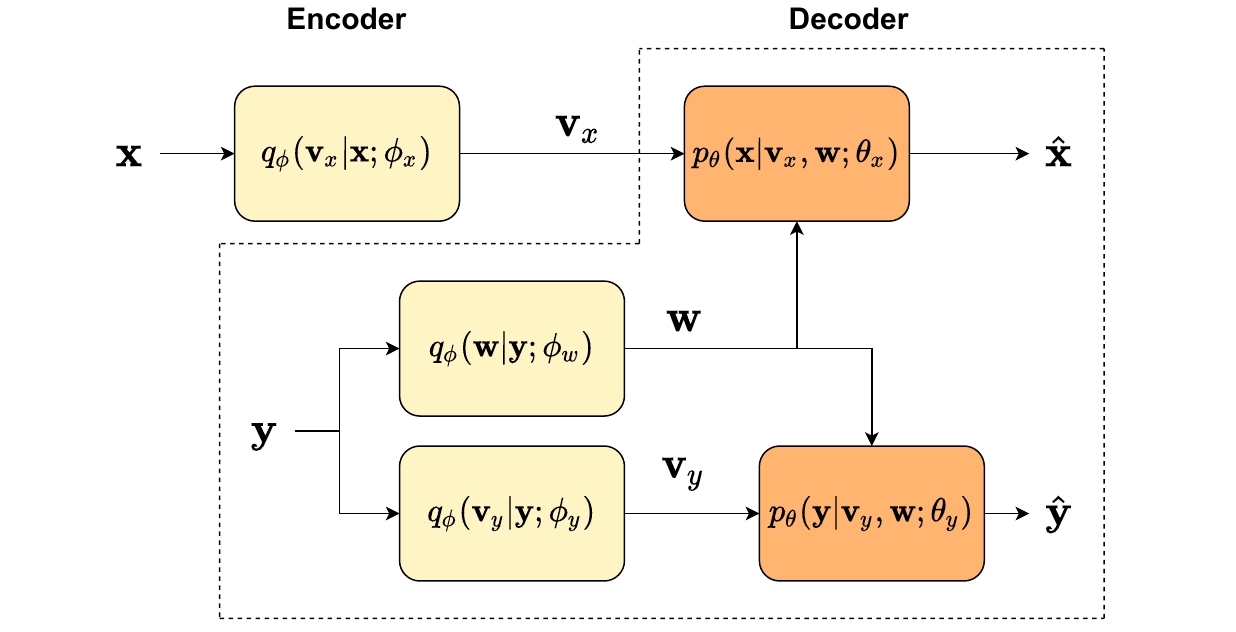} 
\caption{DSC architecture.}
\label{fig:dsc_architecture}
\end{minipage}
\end{figure}

\section{Image compression with side information}\label{wyner_info}

In this section, we describe our main contribution, where we propose a novel autoencoder architecture for image compression exploiting the side information, $\mathbf{y}$, available only at the decoder, to reconstruct the image $\mathbf{x}$. In the Wyner-Ziv characterization of the rate-distortion function, the encoder identifies and transmits the latent variable $V$, which is then combined with the side information to reconstruct the original image. This is the approach followed by \cite{DSIN}. In our architecture, we model the images $\mathbf{x}$ and $\mathbf{y}$ as being generated by the random variables $\mathbf{w}$, $\mathbf{v}_x$ and $\mathbf{v}_y$, according to the graphical model illustrated in Fig.~\ref{fig:prob_model}, which satisfy the Markov chains $\mathbf{v}_x-\mathbf{x}-\mathbf{y}$, $\mathbf{x}-\mathbf{w}-\mathbf{y}$ and $\mathbf{x}-\mathbf{y}-\mathbf{v}_y$. The variable $\mathbf{w}$ captures the common features between the two images, while the variables $\mathbf{v}_x$ and $\mathbf{v}_y$, which we call the private information variables for the respective images, capture those aspects of $\mathbf{x}$ and $\mathbf{y}$ that are not captured by the common variable $\mathbf{w}$. 
We expect that feeding only the common information to the combining function at the decoder, instead of all the side information, will help the encoder to identify and send only the information that is not available at the decoder through the side information, and simplify the task of the combining function.

In general, we have $R^{WZ}_{X|Y}(d) \geq R^{WZ}_{X|W}(d)$; that is, it is more beneficial to have $W$ as side information at the decoder, as it is `closer' to $X$ in distribution. Of course, we cannot in general generate $W$ reliably based only on $Y$, but we argue that even an approximate reconstruction of $W$ allows the decoder to extract only the most relevant parts of the side information that are helpful in estimating $X$. Our experimental results confirm that the proposed approach helps in improving the reconstructed image quality.

\textbf{Architecture: }See Fig.~\ref{fig:dsc_architecture} for the conceptual architecture of the distributed source encoder/decoder pair that we consider, and Fig.~\ref{fig:dsc_architecture2} for the implemented network architecture. The distribution $q_{\boldsymbol{\phi}}(\mathbf{v}_x \mid \mathbf{x};\boldsymbol{\phi}_x)$ is learnt by a transform $\mathbf{g}_{ax}$ at the encoder, and $q_{\boldsymbol{\phi}}(\mathbf{v}_y \mid \mathbf{y};\boldsymbol{\phi}_y)$ is learnt by a transform $\mathbf{g}_{ay}$ at the decoder. The encoder maps the image $\mathbf{x}$ to a latent representation $\mathbf{v}_x$ by applying the transform $\mathbf{g}_{ax}$ to it. The latent representation $\mathbf{v}_x$ is quantized to $\hat{\mathbf{v}}_x \in \mathbb{Z}^m$. Since the quantization step is a non-differentiable operation, which prevents end-to-end training, it is instead replaced by additive uniform random noise over $[-0.5,0.5]$ during training (see \cite{balle2017}). Thus, $\mathbf{v}_x$ is perturbed by uniform noise during training to obtain $\mathbf{\tilde{v}}_{x}$, which approximates the quantized latents $\hat{\mathbf{v}}_x$. The decoder extracts the common information $\mathbf{w} = \mathbf{f}(\mathbf{y};\boldsymbol{\phi}_f)$ between the images $\mathbf{x}$ and $\mathbf{y}$ by applying the transform $\mathbf{f}$, where $\boldsymbol{\phi}_f$ refers to the weights of the DNN. The transform $\mathbf{f}$ learns the marginal distribution $q_{\boldsymbol{\phi}}(\mathbf{w}\mid \mathbf{y};\boldsymbol{\phi}_f)$. The decoder concatenates $\mathbf{w}$ to the received latent variable $\hat{\mathbf{v}}_x$, and reconstructs an estimate $\hat{\mathbf{x}} = \mathbf{g}_{sx}(\hat{\mathbf{v}}_x, \mathbf{w}; \boldsymbol{\theta}_{x})$ of the image $\mathbf{x}$ by applying a transform $\mathbf{g}_{sx}$, which corresponds to the marginal decoder $p_{\boldsymbol{\theta}}(\mathbf{x}\mid \mathbf{v}_x, \mathbf{w};\boldsymbol{\theta}_x)$. Simultaneously, the decoder learns to reconstruct the correlated image $\mathbf{y}$ by first mapping it to the latent representation $\mathbf{v}_y$ using a transform $\mathbf{g}_{ay}$, and then reconstructing an estimate $\hat{\mathbf{y}} = \mathbf{g}_{sy}(\mathbf{v}_y, \mathbf{w}; \boldsymbol{\theta}_y)$ by concatenating the common variable $\mathbf{w}$ to $\mathbf{v}_y$ and applying a transform $\mathbf{g}_{sy}$ to it. Note that the latent representation $\mathbf{v}_y$ is neither quantized nor perturbed with uniform noise. This is because the encoding and decoding of image $\mathbf{y}$ happen inside the decoder without it being transmitted over the channel.

The variables $\mathbf{w}$, $\mathbf{v}_x$ and $\mathbf{v}_y$ are modeled using a univariate non-parametric, fully factorized density function, similarly to the method proposed in \cite{balle2017, balle2018}. The joint distribution of the random variables, assuming the model illustrated in Fig.~\ref{fig:prob_model}, is given by:
\begin{equation}
    p(\mathbf{x},\mathbf{y},\mathbf{w},\mathbf{v}_x,\mathbf{v}_y) = p(\mathbf{w})p(\mathbf{v}_x)p(\mathbf{v}_y)p_{\boldsymbol{\theta}}(\mathbf{x}\mid \mathbf{w},\mathbf{v}_x ; \boldsymbol{\theta}_x)p_{\boldsymbol{\theta}}(\mathbf{y}\mid \mathbf{w},\mathbf{v}_y ; \boldsymbol{\theta}_y),
\end{equation}
parameterized by $\boldsymbol{\theta}_x$ and $\boldsymbol{\theta}_y$. To obtain tractable inference of latent variables, we introduce the following factored variational approximation of the posterior distribution:

\begin{equation}
    q_{\boldsymbol{\phi}}(\mathbf{w},\mathbf{v}_x,\mathbf{v}_y\mid \mathbf{x},\mathbf{y}) = q_{\boldsymbol{\phi}}(\mathbf{v}_x\mid \mathbf{x}; \boldsymbol{\phi}_x)q_{\boldsymbol{\phi}}(\mathbf{w}\mid \mathbf{y}; \boldsymbol{\phi}_w)q_{\boldsymbol{\phi}}(\mathbf{v}_y\mid \mathbf{y}; \boldsymbol{\phi}_y), 
\end{equation}
each of which is parameterized by a distinct DNN. The joint distribution of the latent variables can be estimated by minimizing the expectation of the Kullback-Liebler (KL) divergence between the approximate variational density $q_{\boldsymbol{\phi}}(\tilde{\mathbf{v}}_{x}, \mathbf{v}_{y}, \mathbf{w} \mid \mathbf{x},\mathbf{y})$ and the true posterior distribution $p(\tilde{\mathbf{v}}_{x}, \mathbf{v}_{y}, \mathbf{w}\mid \mathbf{x},\mathbf{y})$ over the data distribution $p(\mathbf{x},\mathbf{y})$:
\begin{align}
    & \mathbb{E}_{\mathbf{x},\mathbf{y} \sim p(\mathbf{x},\mathbf{y})} D_{\mathrm{KL}} \left[q_{\boldsymbol{\phi}}(\tilde{\mathbf{v}}_{x}, \mathbf{v}_{y}, \mathbf{w} \mid \mathbf{x},\mathbf{y}) \mid \mid  p(\tilde{\mathbf{v}}_{x}, \mathbf{v}_{y}, \mathbf{w}\mid \mathbf{x},\mathbf{y})\right] = \mathbb{E}_{\mathbf{x},\mathbf{y} \sim p(\mathbf{x},\mathbf{y})} \mathbb{E}_{\tilde{\mathbf{v}}_{x},\mathbf{v}_{y},\mathbf{w} \sim q_{\boldsymbol{\phi}}}  \Bigg( \nonumber \\
   & \Big( \log q_{\boldsymbol{\phi}}(\tilde{\mathbf{v}}_{x} \mid \mathbf{x}; \boldsymbol{\phi}_x) +  \log q_{\boldsymbol{\phi}}(\mathbf{v}_{y} \mid \mathbf{y}; \boldsymbol{\phi}_y) + \log q_{\boldsymbol{\phi}}(\mathbf{w} \mid \mathbf{y}; \boldsymbol{\phi}_f) \Big) - \Big( \underbrace{\log p_{\boldsymbol{\theta}}(\mathbf{x} \mid \mathbf{w}, \tilde{\mathbf{v}}_{x}; \boldsymbol{\theta}_x)}_{D_x}  \label{variational_loss} \nonumber \\
    &+  \underbrace{\log p_{\boldsymbol{\theta}}(\mathbf{y} \mid \mathbf{w}, \mathbf{v}_{y}; \boldsymbol{\theta}_y)}_{D_y} + \underbrace{\log p(\mathbf{w})}_{R_w} + \underbrace{  \log p(\tilde{\mathbf{v}}_{x})}_{R_x} + \underbrace{\log p(\mathbf{v}_{y})}_{R_y} \Big) \Bigg) + \text{const.}, 
\end{align}
where the first term, $q_{\boldsymbol{\phi}}(\tilde{\mathbf{v}}_{x} \mid \mathbf{x}; \boldsymbol{\phi}_x)$, is a constant because the quantization noise has a uniform distribution with a constant width, and the second and third terms are zero because the inference model is deterministic for the common information and the latent representation of the correlated image, which implies that the associated conditional entropies are zero. If we assume a squared-error distortion metric, the terms $D_x$ and $D_y$ correspond to weighted distortion terms for the reconstruction of the input image and the correlated image respectively. The entropy terms $R_w$, $R_x$ and $R_y$ correspond to `rate-like' terms for the common information, the quantized latent representation of the input image, and the latent representation of the correlated image, respectively. For perceptual distortion metrics, the terms in Eq.~\ref{variational_loss} may not correspond to the rate-distortion interpretation. However, we retain the same form of the loss function, and train the DNN by minimizing the following loss function:
\begin{align}
    L(\mathbf{g}_{ax},\mathbf{g}_{sx},\mathbf{g}_{ay},\mathbf{g}_{sy},\mathbf{f}) = \left( R_x + \lambda D_x \right) + \alpha \left(R_y + \lambda D_y\right) + \beta R_w,\label{loss_full}
\end{align}
where, for simplicity, we use the same weight $\lambda$ for the distortion terms. Since our main objective is only to reconstruct $\mathbf{x}$, we introduce the hyperparameters $\alpha$ and $\beta$ that determine how much importance is given to the reconstruction of the correlated image, and to the complexity of common information extracted by the decoder, respectively, which act as regularizers for the main objective. We also extend the above solution to use scale hyperpriors by using the model of \cite{balle2018} as the baseline.

\section{Experiments} \label{sec:experiments}
To compare the compression performance of our proposed model with the state-of-the-art, we conducted a number of experiments using the PyTorch framework. Our code is publicly available\footnote{Our code is available at: \url{https://github.com/ipc-lab/NDIC}}.

\textbf{Experimental setup: }Fig.~\ref{fig:dsc_architecture2} illustrates the proposed DNN architecture for distributed source coding in detail. The transforms $\mathbf{g}_{ax}$ and $\mathbf{g}_{sx}$ have the same structure as those in \cite{balle2017}. These transforms are composed of convolutional layers, and linear (i.e., rectified linear unit) and nonlinear functions (i.e., generalized divisive normalization [GDN] and inverse generalized divisive normalization [IGDN]), which have been shown to be particularly suitable for density modelling in image compression \cite{balle2017}. Note that, we have omitted the section of the network that the decoder uses to reconstruct the correlated image during training in Fig.~\ref{fig:dsc_architecture2}. We introduce the transform $\mathbf{f}$ as mentioned in Section~\ref{wyner_info}. 

For the first set of experiments, we constructed our dataset from the KITTI Stereo 2012 dataset  \cite{Geiger2012CVPR} and from the KITTI Stereo 2015 dataset \cite{Menze2015ISA,Menze2018JPRS}, consisting of unique 1578 stereo image pairs (i.e., a pair of two images taken simultaneously by different cameras). We refer to this dataset as \emph{KITTI Stereo}. We trained every model on 1576 image pairs, and we validated and tested every model on 790 image pairs from KITTI Stereo dataset. We also trained our models on the \emph{Cityscape} dataset \cite{Cordts2016Cityscapes}, consisting of 5000 stereo image pairs, out of which 2975 image pairs belong to the training dataset, 500 image pairs belong to the validation dataset, and 1525 image pairs belong to the test dataset. These datasets are designed to illustrate the calibrated and synchronized camera array use case.

\textbf{Evaluation: }We evaluate our models using both the multi-scale structural similarity index measure (MS-SSIM), as well as the peak SNR (PSNR) based on the mean-squared error (MSE) distortion. The MS-SSIM has been widely reported to provide a better measure of human perception of distortion \cite{MS-SSIM}.

\textbf{Training: }We center-crop each $375 \times 1242$ image of the KITTI Stereo dataset to obtain a $370 \times 740$ image, and then downsample it to a $128 \times 256$ image. For Cityscape dataset, we directly downsample each image to $128 \times 256$. We train the baseline model with different values of $\lambda$ to obtain points with different bit rates using the PSNR and MS-SSIM metrics for the reconstruction loss. We train the proposed model for 500K iterations, using randomly initialized network weights. Similarly to \cite{balle2017}, we train our models using AMSGrad optimizer \cite{j.2018on}, with a learning rate of $1 \times 10^{-4}$. The learning rate is reduced by a factor of $10$ whenever the decrease in the loss function stagnates, where the lower bound on the learning rate is set to $1 \cdot 10^{-7}$. We use a batch size of 1 because of the small size of the datasets under consideration. For comparison, we also train the model proposed in \cite{DSIN} by using the provided code\footnote{\url{https://github.com/ayziksha/DSIN}}, which will be referred to as DSIN. We highlight that the results for DSIN reported here differ from those in \cite{DSIN} because we use smaller images in our experiments.

\begin{figure}
    \begin{subfigure}{0.49\textwidth}
    \centering
    \includegraphics[width=1\textwidth]{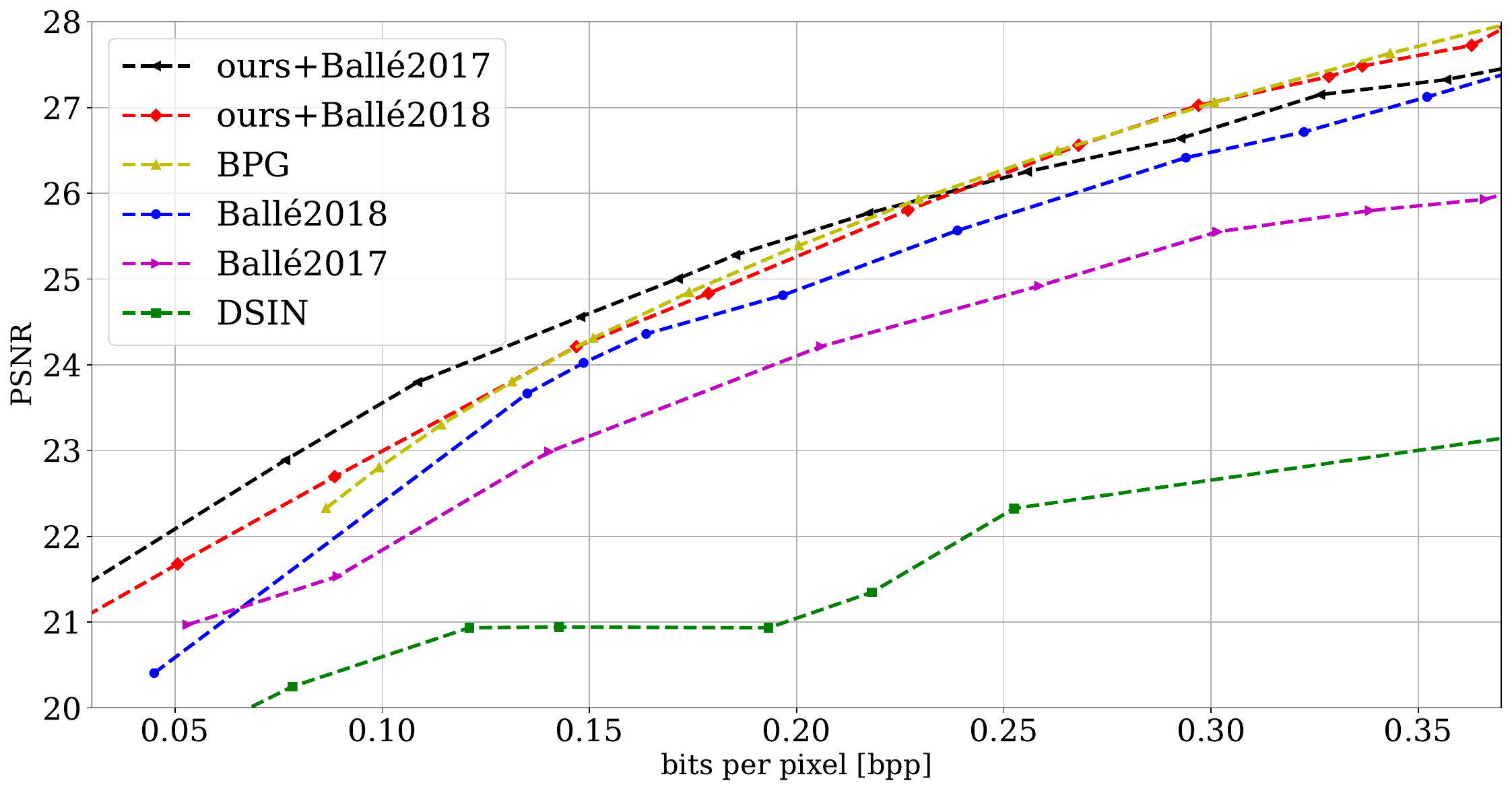}
    \caption{KITTI Stereo (PSNR)}
    \label{fig:psnr}
    \end{subfigure}%
   \centering
   \hfill
    \begin{subfigure}{0.49\textwidth}
     \centering
      \includegraphics[width=1\textwidth]{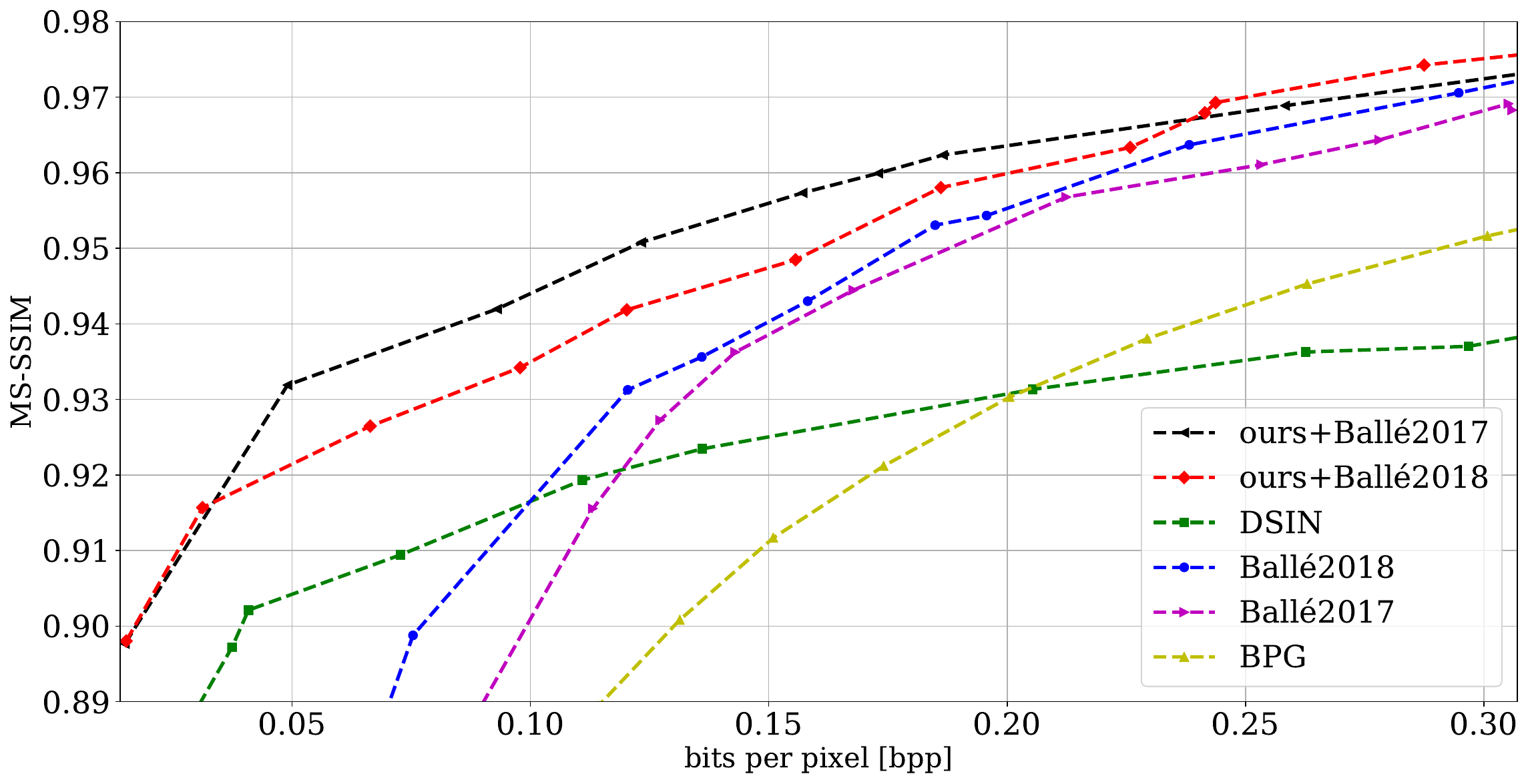}
      \caption{KITTI Stereo (MS-SSIM)}
    \label{fig:msssim}
    \end{subfigure}%
    \\
    \begin{subfigure}{0.49\textwidth}
    \centering
    \includegraphics[width=1\textwidth]{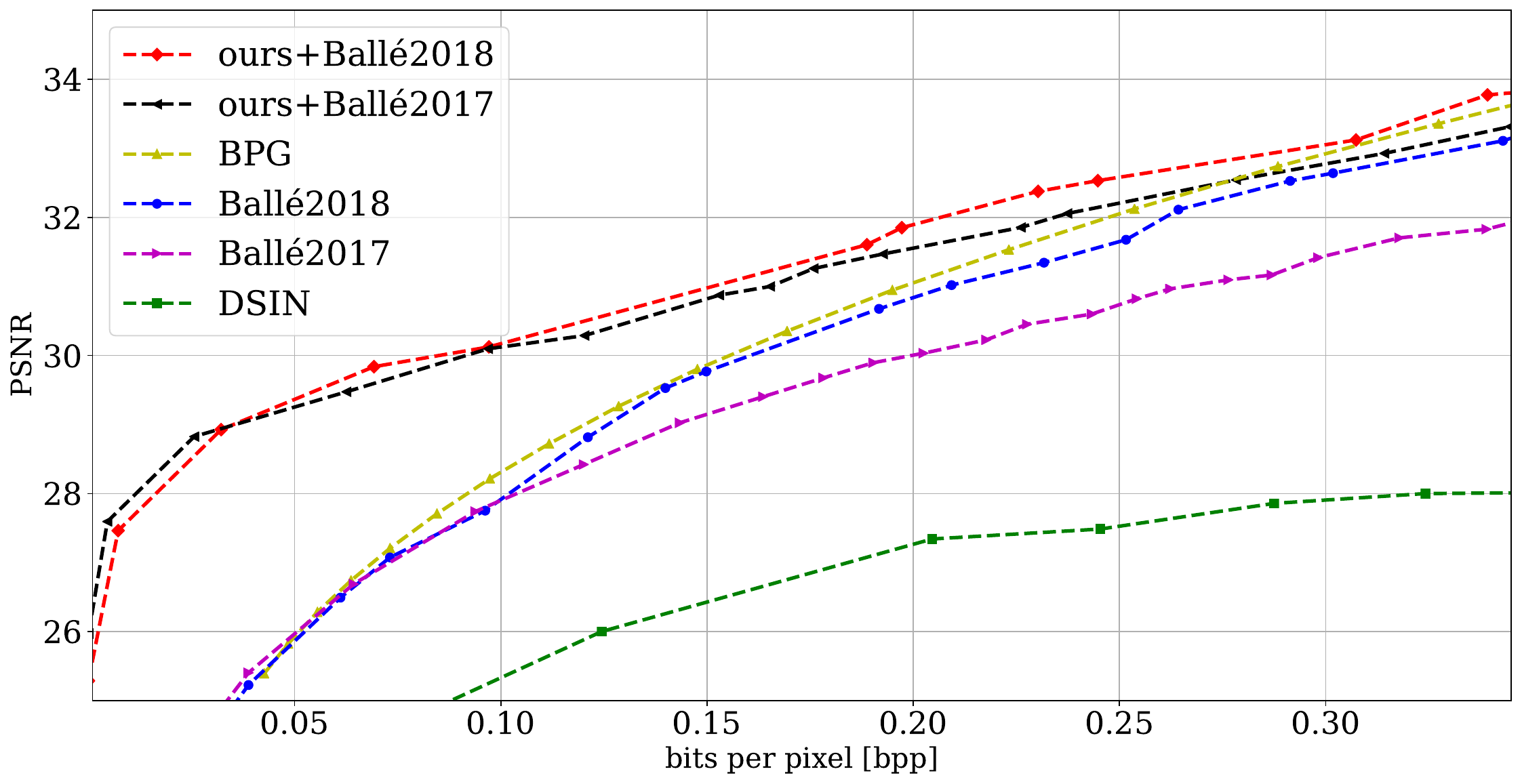}
    \caption{Cityscape (PSNR)}
    \label{fig:cs_psnr}
    \end{subfigure}%
    \centering
    \hfill
    \begin{subfigure}{0.49\textwidth}
    \includegraphics[width=1\textwidth]{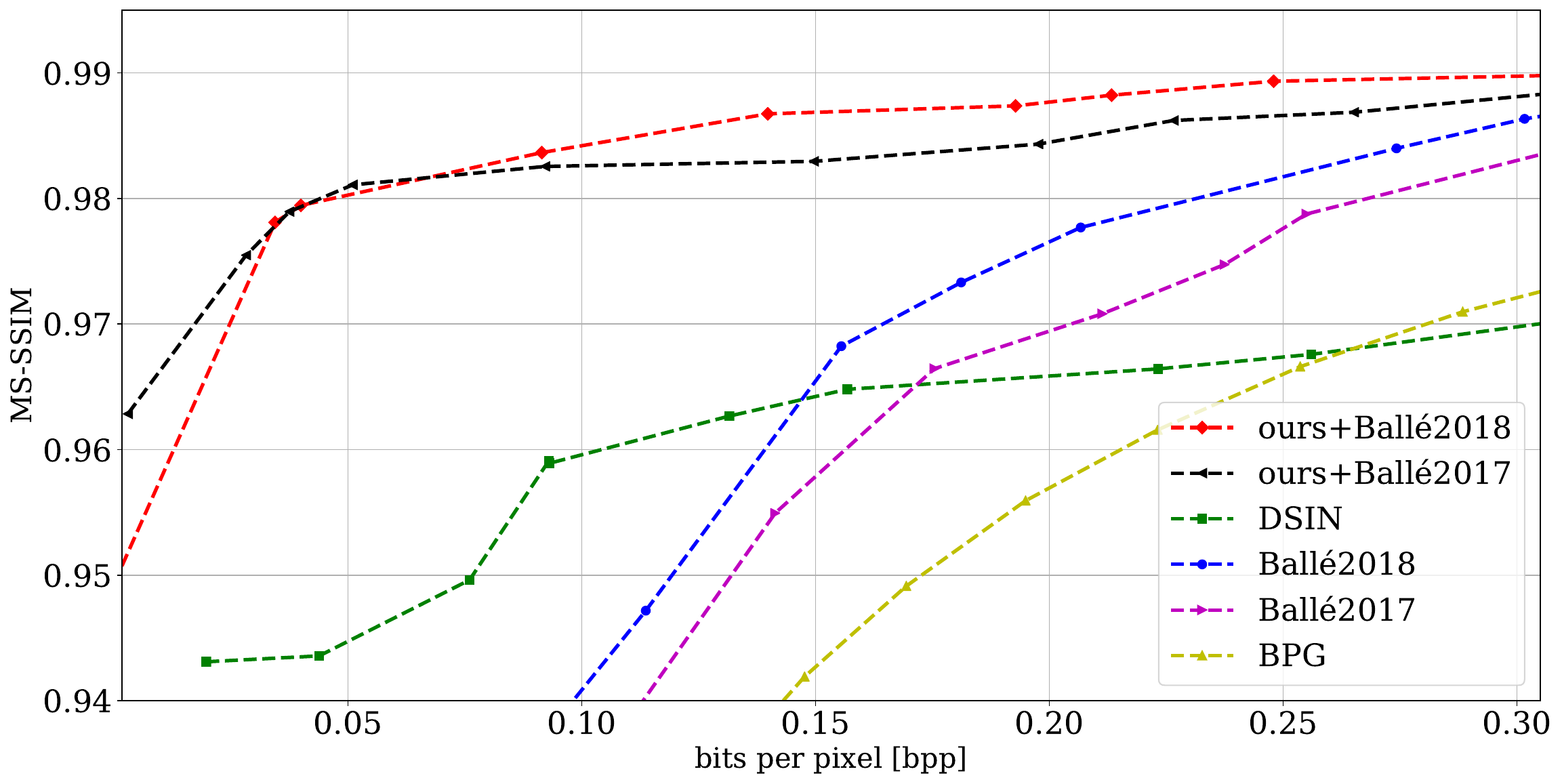}
    \caption{Cityscape (MS-SSIM)}
    \label{fig:cs_msssim}
    \end{subfigure}%
    \caption{Comparison of different models in terms of MSE and MS-SSIM metrics. }
    \label{fig:comparisons_plot}
\end{figure}

\textbf{Experimental results: }In this section, we discuss the performance of the proposed model and compare it with other alternatives (see Fig.~\ref{fig:comparisons_plot}). In addition to DSIN, we also consider BPG as well as DNN-aided compression schemes introduced in \cite{balle2017} and \cite{balle2018}, which will be referred to as Ballé2017 and Ballé2018, respectively. Following \cite{BPG_chroma}, we employ 4:4:4 chroma format for BPG. We note that these schemes do not exploit the side information at the receiver. ``ours + Ballé2017'' and ``ours + Ballé2018'' refer to our proposed solutions with the models in \cite{balle2017} and \cite{balle2018} as the baselines, respectively. In Figs.~\ref{fig:psnr} and \ref{fig:cs_psnr}, we present the comparison in terms of the average PSNR. We observe that the proposed model is particularly effective at low bit rates. Note that DSIN does not perform well when optimized for MSE. We observe an even more stark improvement in performance when optimized with respect to MS-SSIM in Fig.~\ref{fig:msssim} and \ref{fig:cs_msssim}, where the models that do not use side information experience a sharp drop in performance at low bit rates, while DSIN and the proposed solution, which exploit the side information, experience a more graceful degradation of performance with decreasing bit rates. This illustrates the fact that the presence of side information is particularly useful at low bit rates, where even a very limited information allows reasonable reconstruction by exploiting the side information. We also note that the proposed solution significantly improves the performance with respect to DSIN in experiments with both datasets. 

\begin{figure}
    \centering
    \hspace*{-0.8cm} 
    \begin{tabular}[b]{cccc}
     Original Image & Ballé2018 & DSIN & Ours
     \\
      \begin{subfigure}[t]{0.25\textwidth}
          \centering
          \includegraphics[width=0.95\linewidth]{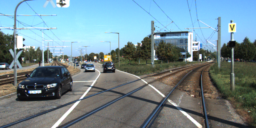}
          \caption{KITTI Stereo}
          \label{fig:sub10}
       \end{subfigure}&
       \begin{subfigure}[t]{0.25\textwidth}
          \centering
          \includegraphics[width=0.95\linewidth]{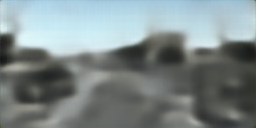}
          \caption{bpp = 0.0261}
          \label{fig:sub1}
       \end{subfigure}&
       \begin{subfigure}[t]{0.25\textwidth}
          \centering
          \includegraphics[width=0.95\linewidth]{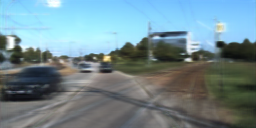}
          \caption{bpp = 0.0187}
          \label{fig:sub2}
       \end{subfigure}&
       \begin{subfigure}[t]{0.25\textwidth}
          \centering
          \includegraphics[width=0.95\linewidth]{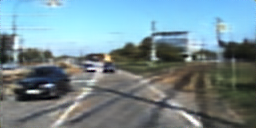}
          \caption{bpp = 0.0152}
          \label{fig:sub3}
       \end{subfigure}\\
       
       \begin{subfigure}[t]{0.25\textwidth}
        	\centering
        	\includegraphics[width=0.95\linewidth]{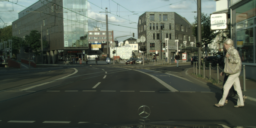}
        	\caption{Cityscape}
        	\label{fig:app0_7}
        \end{subfigure}&
        \begin{subfigure}[t]{0.25\textwidth}
        	\centering
        	\includegraphics[width=0.95\linewidth]{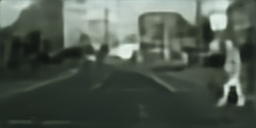}
        	\caption{bpp = 0.0782}
        	\label{fig:app1_7}
        \end{subfigure}&
        \begin{subfigure}[t]{0.25\textwidth}
        	\centering
        	\includegraphics[width=0.95\linewidth]{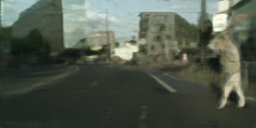}
        	\caption{bpp = 0.0460
        	}
        	\label{fig:app3_7}
        \end{subfigure}&
        \begin{subfigure}[t]{0.25\textwidth}
        	\centering
        	\includegraphics[width=0.95\linewidth]{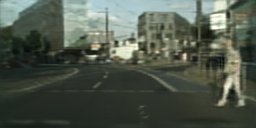}
        	\caption{bpp = 0.0348
        	}
        	\label{fig:app2_7}
        \end{subfigure}

    \end{tabular}
  \caption{Visual comparison of different models trained for the MS-SSIM metric. ``Ours'' in the figures above refers to ``ours + Ballé2017" model.}
    \label{fig:images}
\end{figure}

\begin{figure}
    \hspace*{-0.8cm} 
    \begin{tabular}[b]{cccc}
      \begin{subfigure}[b]{0.25\columnwidth}
        \includegraphics[scale=0.4]{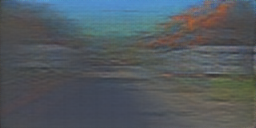}
        \label{fig:sub1_acpr}
      \end{subfigure}&
      \begin{subfigure}[b]{0.25\columnwidth}
        \includegraphics[scale=0.4]{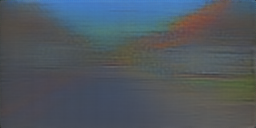}
        \label{fig:sub2_acpr}
      \end{subfigure}&
      \begin{subfigure}[b]{0.25\columnwidth}
        \includegraphics[scale=0.4]{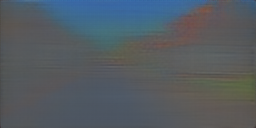}
        \label{fig:sub3_acpr}
      \end{subfigure}&
      \begin{subfigure}[b]{0.25\columnwidth}
        \includegraphics[scale=0.4]{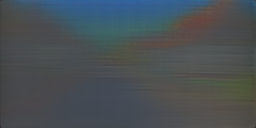}
        \label{fig:sub10_acpr}
      \end{subfigure}\\
      \begin{subfigure}[b]{0.25\columnwidth}
        \includegraphics[scale=0.4]{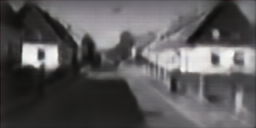}
        \label{fig:sub4_acpr}
      \end{subfigure}&
      \begin{subfigure}[b]{0.25\columnwidth}
        \includegraphics[scale=0.4]{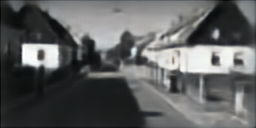}
        \label{fig:sub11_acpr}
      \end{subfigure}&
      \begin{subfigure}[b]{0.25\columnwidth}
        \includegraphics[scale=0.4]{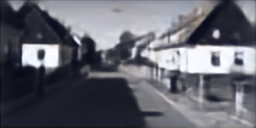}
        \label{fig:sub5_acpr}
      \end{subfigure}&
      \begin{subfigure}[b]{0.25\columnwidth}
        \includegraphics[scale=0.4]{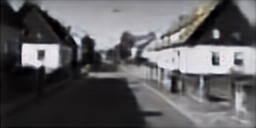}
        \label{fig:sub6_acpr}
      \end{subfigure}\\
      \begin{subfigure}[b]{0.25\columnwidth}
        \includegraphics[scale=0.4]{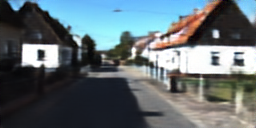}
        \caption{$\alpha=1, \beta=10^{-3}$\\ \centering bpp = 0.1281}
        \label{fig:sub7_acpr}
      \end{subfigure}&
      \begin{subfigure}[b]{0.25\columnwidth}
        \includegraphics[scale=0.4]{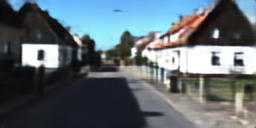}
        \caption{$\alpha=0, \beta=0$ \\ \centering bpp = 0.1572}
        \label{fig:sub9_acpr}
      \end{subfigure}&
      \begin{subfigure}[b]{0.25\columnwidth}
        \includegraphics[scale=0.4]{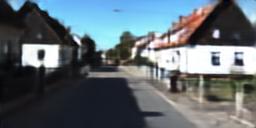}
        \caption{$\alpha=1,\beta=1$ \\ \centering bpp = 0.1658}
        \label{fig:sub8_acpr}
      \end{subfigure}&
      \begin{subfigure}[b]{0.25\columnwidth}
        \includegraphics[scale=0.4]{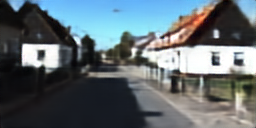}
        \caption{$\alpha=0, \beta=1$ \\ \centering bpp = 0.1731}
        \label{fig:sub12_acpr}
      \end{subfigure}
    \end{tabular}
  \caption{\textbf{Ablation study:} Effect of hyperparameters $\alpha$ and $\beta$ on the common information ($1^{st}$ row) and private information ($2^{nd}$ row) decomposition, for a similar reconstruction quality ($3^{rd}$ row).}
    \label{fig:ablation_common_private}
\end{figure}

We also present a visual comparison of the performance of the models proposed in Fig.~\ref{fig:images}. Notice that the Ballé2018 model fails to capture the colours at very low bit rates, as illustrated in Fig.~\ref{fig:images}, unlike DSIN and ours. Moreover, our model successfully captures the textures and details of colours and objects in the background, while DSIN has a blurring effect on the image. This is because the DSIN model operates by first reconstructing the image based on the compressed description provided by the encoder, then finding the offset of corresponding patches (using the side information finder block in their model) in the intermediate reconstructions of the input and correlated images when passed through their baseline autoencoder, and then using the corresponding patches from the original side information image to refine patches in the intermediate reconstruction of the input image. If the intermediate reconstructions are not of a high quality to begin with, it can cause the wrong patches to be recognized as the corresponding patches in the side information image. This causes distortions in the reconstructed image, leading to low MS-SSIM values. Our model operates instead by overlaying the content details sent by the encoder over the structural and texture details extracted from the side information. This is illustrated in Fig.~\ref{fig:ablation_common_private}. We generate the visualizations of the private and common information by passing them individually through the decoder while setting the other to zero. It is observed that the common information captures the colors and texture information, which explains why our model is able to capture them even at very low bit rates. 
We also note in Fig.~\ref{fig:ablation_common_private} that as the bit rate is increased, the encoder is allowed to send more content and structural information; and therefore, the decoder tends to extract less definite content details and more global color and texture details from the side information.

\begin{figure}[b]
\centering
      \includegraphics[ scale=0.2]{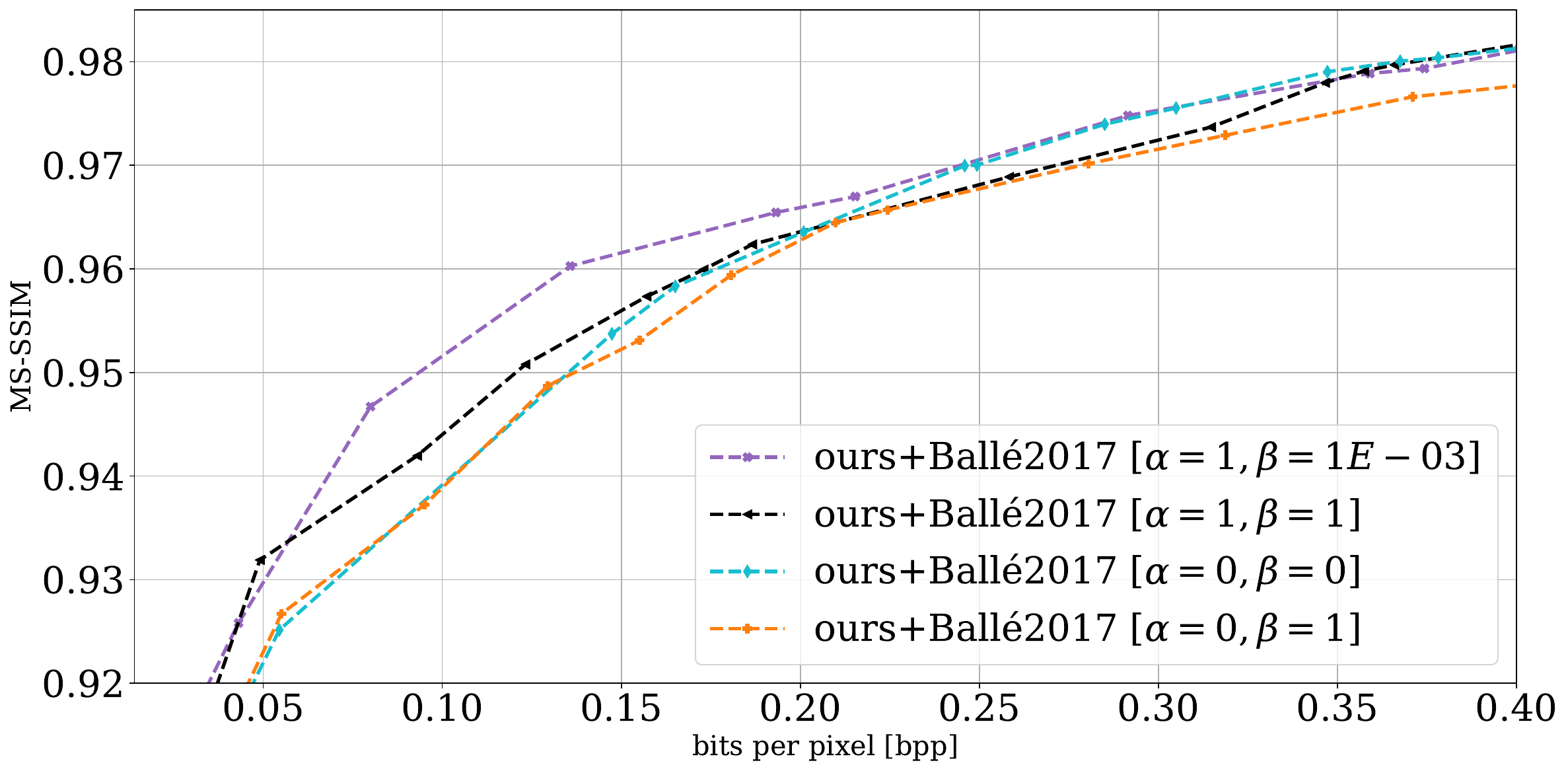}
      \caption{Ablation study experiments for the MS-SSIM metric on KITTI Stereo.}
      \label{fig:ablation_plot}
\end{figure}

\textbf{Ablation study: }To study the impact of each component of the proposed model on the overall performance, we carry out an ablation study on the architecture of the decoder by varying the parameters $\alpha$ and $\beta$ in Eq.~\eqref{loss_full}, and compare the performances in Figs.~\ref{fig:ablation_plot} and \ref{fig:ablation_common_private}. The default model sets $\alpha=\beta = 1$. By setting $\alpha=\beta=0$, we remove the additional regularization terms, and we observe that this results in a slight decrease in the performance at low bit rate. Moreover, we observed an unpredictable behaviour when evaluating the DSIN model, which lacks the regularization terms $R_w$ and $R_y + \lambda D_y$ that we have. Having non-zero $\alpha$ and $\beta$ provides the required regularization, and as a result, improves the performance. We also check whether the desired regularization can be obtained through only the rate of the common part, $W$, by setting $\alpha=0$ and $\beta=1$. However, we observe that the performance is even worse in this setting. We argue that by imposing the decoder to reconstruct its own side information under a rate and distortion penalty prevents $W$ from being too far from $Y$, and hence, in a way, acts as another regularizer on the common part $W$. Making the value of $\beta$ small, with $\alpha=1$, is observed to provide an improvement over the default setting. This is illustrated in Fig.~\ref{fig:ablation_common_private}, where having $\alpha=1$ and $\beta=10^{-3}$ allows the decoder to generate more common information, and therefore, the encoder can send a lower quality private information image at a lower bit rate to achieve the same reconstruction quality. This suggests that $\beta$ can be tuned to maximize the model's performance at different bit rates.

\section{Conclusions} \label{sec:conclusion}
We presented a novel autoencoder for lossy image compression with decoder side information exploiting the common information between the image to be reconstructed and the side information. The encoder learns to send only input image specific information, like the content details, to the decoder, while common information, like texture and colors, are extracted by the decoder from the side information. We show that this approach allows good quality image reconstruction even at very low bit rates, improving significantly over both single image compression models, as well as previous work on image compression with side information. The loss function in Eq.~\eqref{loss_full} also provides a framework to extend this work to a setting where there are two distributed encoders having correlated images. Moreover, combining the side information with the received latent representation in the latent space provides a general framework to extend this work to task-aware image compression.

\Section{References}
\bibliographystyle{IEEEbib}
\bibliography{main}

\section{Appendix}

\subsection{Implementation details - Ballé2017 baseline}

Training each model took around  $\sim36$ hours per model, i.e., obtaining one point on the curves plotted in Fig.~\ref{fig:psnr} and \ref{fig:msssim}. As shown in Fig.~\ref{fig:dsc_architecture2}, our model consists of three main blocks $\mathbf{g}_{ax}$ for encoding the input image to a latent vector, $\mathbf{f}$ for extracting the  common information $\mathbf{w}$ from the side information, and $\mathbf{g}_{sx}$ for decoding the concatenation of common information and the quantized latent vector to the reconstructed image. In Table~\ref{tab:architecture1} and \ref{tab:architecture2}, a top-down architecture of each block is provided. Convolution layer parameters are denoted as: $\text{number of filter} \times \text{ kernel height } \times \text{ kernel width } / \text{ sampling stride} $, where $\uparrow$ and $\downarrow$ indicate upsampling and downsampling, respectively. GDN (IGDN) corresponds to (inverse) generalized divisive normalization operation described in \cite{balle2017}.

\begin{table}[H]
\centering
\caption{Network architecture for $\mathbf{g}_{ax}$ and $\mathbf{f}$ blocks.}
\begin{tabular}{c}
Layers                                   \\ \hline
Conv2D ($192\times 5\times 5\ /2\downarrow$) \\
GDN                                      \\
Conv2D ($192\times 5\times 5\ /2\downarrow$) \\
GDN                                      \\
Conv2D ($192\times 5\times 5\ /2\downarrow$) \\
GDN                                      \\
Conv2D ($192\times 5\times 5\ /2\downarrow$)
\end{tabular}
\label{tab:architecture1}
\hfill
\caption{Network architecture for $\mathbf{g}_{sx}$ block.}
\begin{tabular}{c}
Layers                                   \\ \hline
Conv2D ($192\times 5\times 5\ /2\uparrow$) \\
IGDN                                      \\
Conv2D ($192\times 5\times 5\ /2\uparrow$) \\
IGDN                                      \\
Conv2D ($192\times 5\times 5\ /2\uparrow$) \\
IGDN                                      \\
Conv2D ($3\times 5\times 5\ /2\uparrow$)
\end{tabular}
\label{tab:architecture2}
\end{table}

During training, to simulate quantization of the latent representation and to compute the likelihoods of the latent representation, an entropy bottleneck block is used, which corresponds to the univariate non-parametric density model used for modeling the distribution of the latent representation. The entropy bottleneck block implementation is similar to the one provided in \cite{balle2018}. The entropy bottleneck block is also used to estimate the likelihoods of the common variable $\mathbf{w}$.

During training, the side information image $\mathbf{y}$ is also passed through the autoencoder, where it is first mapped to its latent representation using the encoder block $\mathbf{g}_{ay}$, which has the same structure as $\mathbf{g}_{ax}$, and then reconstructing $\mathbf{y}$ by passing its latent representation through the decoder block $\mathbf{g}_{sy}$, which has the same structure as $\mathbf{g}_{sx}$. Note that the latent representation of $\mathbf{y}$ is not quantized.

Note that we opt for sliding Gaussian window of size $7 \times 7$, instead of the popular choice of $11 \times 11$, for MS-SSIM reconstruction loss function calculations due to having utilised image size of $128 \times 256$ in our training setup.

\subsection{Wyner-Ziv rate-distortion function with Wyner common information}\label{app:Wyner}

Let $W$ denote the Wyner common information between source $X$ and side information $Y$, where $X-W-Y$ form a Markov chain. It is clear that the rate-distortion function would reduce if we provide $W$ as additional side information to the receiver, i.e., we have $R^{WZ}_{X \mid Y}(d) \geq R^{WZ}_{X\mid YW}(d)$, $\forall d\geq 0$. On the other hand, note that, $R^{WZ}_{X\mid YW}(d) = \inf I(X;V \mid W, Y)$, over all $V$ for which $V-X-W-Y$ form a Markov chain, and there exist a function $f$ that satisfies $\mathbb{E}\left[ D(X,f(V,W,Y)) \right] \leq d$. Since, $I(X;V \mid W, Y) = I(X;V \mid W)$, and conditioned on $W$, $Y$ does not help in the estimation of $X$. Hence, we conclude that $R^{WZ}_{X \mid Y}(d) \geq R^{WZ}_{X \mid YW}(d) = R^{WZ}_{X \mid W}(d)$.

\subsection{Derivation of the variational loss function in Eq.~\eqref{variational_loss}}
We have
\begin{align}
 & \mathbb{E}_{\mathbf{x},\mathbf{y} \sim p(\mathbf{x},\mathbf{y})} D_{\mathrm{KL}} \left[q_{\boldsymbol{\phi}}(\tilde{\mathbf{v}}_{x}, \mathbf{v}_{y}, \mathbf{w} \mid \mathbf{x},\mathbf{y}) \mid\mid p(\tilde{\mathbf{v}}_{x}, \mathbf{v}_{y}, \mathbf{w}\mid \mathbf{x},\mathbf{y})\right] \nonumber \\
&=  \mathbb{E}_{\mathbf{x},\mathbf{y} \sim p(\mathbf{x},\mathbf{y})} \mathbb{E}_{\tilde{\mathbf{v}}_{x},\mathbf{v}_{y},\mathbf{w} \sim q_{\boldsymbol{\phi}}} \Bigg[ \log\left( \frac{q_{\boldsymbol{\phi}}(\tilde{\mathbf{v}}_{x}, \mathbf{v}_{y}, \mathbf{w} \mid \mathbf{x},\mathbf{y})}{p(\tilde{\mathbf{v}}_{x}, \mathbf{v}_{y}, \mathbf{w}\mid \mathbf{x},\mathbf{y})} \right) \Bigg] \\ 
&=  \mathbb{E}_{\mathbf{x},\mathbf{y} \sim p(\mathbf{x},\mathbf{y})} \mathbb{E}_{\tilde{\mathbf{v}}_{x},\mathbf{v}_{y},\mathbf{w} \sim q_{\boldsymbol{\phi}}} \Bigg[ \log\left( \frac{q_{\boldsymbol{\phi}}(\tilde{\mathbf{v}}_{x}, \mathbf{v}_{y}, \mathbf{w} \mid \mathbf{x},\mathbf{y}) p(\mathbf{x},\mathbf{y})}{p(\tilde{\mathbf{v}}_{x}, \mathbf{v}_{y}, \mathbf{w}, \mathbf{x},\mathbf{y})} \right) \Bigg]\\
&=  \mathbb{E}_{\mathbf{x},\mathbf{y} \sim p(\mathbf{x},\mathbf{y})} \mathbb{E}_{\tilde{\mathbf{v}}_{x},\mathbf{v}_{y},\mathbf{w} \sim q_{\boldsymbol{\phi}}} \Bigg[ \log\left( \frac{q_{\boldsymbol{\phi}}(\tilde{\mathbf{v}}_{x}\mid \mathbf{x};\boldsymbol{\phi}_x) q_{\boldsymbol{\phi}}(\mathbf{v}_{y} \mid \mathbf{y};\boldsymbol{\phi}_y) q_{\boldsymbol{\phi}}(\mathbf{w} \mid \mathbf{y};\boldsymbol{\phi}_f) p(\mathbf{x},\mathbf{y})}{p(\mathbf{x} \mid \tilde{\mathbf{v}}_{x}, \mathbf{w};\boldsymbol{\theta}_x) p(\mathbf{y} \mid \mathbf{v}_{y}, \mathbf{w};\boldsymbol{\theta}_y) p(\mathbf{w}) p(\tilde{\mathbf{v}}_x) p(\mathbf{v}_y)} \right) \Bigg] \label{factorization} \\
& = \mathbb{E}_{\mathbf{x},\mathbf{y} \sim p(\mathbf{x},\mathbf{y})} \mathbb{E}_{\tilde{\mathbf{v}}_{x},\mathbf{v}_{y},\mathbf{w} \sim q_{\boldsymbol{\phi}}}  \Bigg( \Big( \log q_{\boldsymbol{\phi}}(\tilde{\mathbf{v}}_{x} \mid \mathbf{x}; \boldsymbol{\phi}_x) +  \log q_{\boldsymbol{\phi}}(\mathbf{v}_{y} \mid \mathbf{y}; \boldsymbol{\phi}_y) + \log q_{\boldsymbol{\phi}}(\mathbf{w} \mid \mathbf{y}; \boldsymbol{\phi}_f) \Big) \nonumber \\
   & - \Big( \underbrace{\log p_{\boldsymbol{\theta}}(\mathbf{x} \mid \mathbf{w}, \tilde{\mathbf{v}}_{x}; \boldsymbol{\theta}_x)}_{D_x} 
    +  \underbrace{\log p_{\boldsymbol{\theta}}(\mathbf{y} \mid \mathbf{w}, \mathbf{v}_{y}; \boldsymbol{\theta}_y)}_{D_y} + \underbrace{\log p(\mathbf{w})}_{R_w} + \underbrace{  \log p(\tilde{\mathbf{v}}_{x})}_{R_x} + \underbrace{\log p(\mathbf{v}_{y})}_{R_y} \Big) \Bigg)+ const.
\end{align}

\subsection{Hyperprior-based model extension}
We extend the proposed model to include the hyperpriors $\mathbf{z}_x$ and $\mathbf{z}_y$, similarly to \cite{balle2018}, by stacking the parametric transforms $\mathbf{h}_{ax}$ and $\mathbf{h}_{ay}$ on top of $\mathbf{v}_x$ and $\mathbf{v}_y$, respectively. The hyperprior models the spatial dependencies between the elements of the latent variable, and enables better compression by the entropy coder. Conditioned on the hyperprior variable, each element of $\mathbf{v}_x$, denoted by $v_x(i), i=1,\ldots, m$, is assumed to be an independent zero-mean Gaussian with its own standard deviation $\sigma_i$, where the standard deviations are predicted by applying a parametric transform $\mathbf{h}_{sx}$ to the hyperprior $\hat{\mathbf{z}}_x$.

Similarly to \cite{balle2018}, the quantization of the hyperprior is replaced by perturbing it with uniform random noise during training to obtain $\tilde{\mathbf{z}}_x$. The joint density of $\tilde{\mathbf{v}}_x$ and $\tilde{\mathbf{z}}_x$ through the inference mechanism is modeled as:
\begin{align}
        q_{\boldsymbol{\phi}}(\tilde{\mathbf{v}}_x, \tilde{\mathbf{z}}_x \mid \mathbf{x};\phi_x ) &= \prod_{i} \mathcal{U}\left( \tilde{v}_x(i) \Bigm| v_x(i) -\frac{1}{2}, v_x(i) + \frac{1}{2} \right) \cdot \prod_{j} \mathcal{U}\left( \tilde{z}_x(j) \Bigm| z_x(j) -\frac{1}{2}, z_x(j) + \frac{1}{2} \right).
\end{align}
The prior of $\tilde{\mathbf{v}}_x$, conditioned on the perturbed hyperprior $\tilde{\mathbf{z}}_x$, is given by:
\begin{align}
    p(\tilde{\mathbf{v}}_x \mid \tilde{\mathbf{z}}_x) &= \prod_{i} \left( \mathcal{N}(0,\tilde{\sigma}_i^2) * \mathcal{U}(-0.5,0.5) \right) \left(\tilde{v}_x(i)\right)\\
    & \text{with } \tilde{\sigma}_i = \mathbf{h}_s(\tilde{\mathbf{z}}_x;\boldsymbol{\theta}_h),
\end{align}
where $\boldsymbol{\theta}_h$ refers to the weight of the neurons, and $\mathcal{N}(\cdot, \cdot)$ and $\mathcal{U}(\cdot, \cdot)$ correspond to the normal distribution and the uniform distribution on $\tilde{v}_x(i)$. As we assume no prior beliefs about the hyperpriors, the probability density of the perturbed hyperpriors $\tilde{\mathbf{z}}_x$ is modeled using a univariate non-parametric, fully factorized density model \cite{balle2018}:
\begin{align}
    p(\tilde{\mathbf{z}}_x) = \prod_{i} \left( p_{z_x(i)\mid \text{ } \boldsymbol{\psi}^{(i)}}( \boldsymbol{\psi}^{(i)} ) * \mathcal{U}(-0.5, 0.5) \right) (z_x(i)),
\end{align}
where the vectors $\boldsymbol{\psi}^{(i)}$ encapsulate the parameters of each univariate distribution $p_{z_x(i)\mid \text{ } \boldsymbol{\psi}^{(i)}}$. During evaluation, the quantized latent representation $\tilde{\mathbf{v}}_x$ and the quantized hyperprior $\tilde{\mathbf{z}}_x$ are encoded and transmitted as a bit-stream by the arithmetic encoder using the Gaussian prior density model and the univariate non-parametric density model respectively.

Employing the method of variational inference, the loss function of this model works out to be
\begin{align}
   & \mathbb{E}_{\mathbf{x},\mathbf{y} \sim p(\mathbf{x},\mathbf{y})} \mathrm{D}_{\mathrm{KL}} \left[q_{\boldsymbol{\phi}}(\tilde{\mathbf{v}}_{x}, \mathbf{v}_{y}, \mathbf{w}, \tilde{\mathbf{z}}_x, \mathbf{z}_y \mid \mathbf{x},\mathbf{y}) \mid \mid  p_{\boldsymbol{\theta}}(\tilde{\mathbf{v}}_{x}, \mathbf{v}_{y}, \mathbf{w},\tilde{\mathbf{z}}_x, \mathbf{z}_y \mid \mathbf{x},\mathbf{y})\right]  \\
   =& \mathbb{E}_{\mathbf{x},\mathbf{y} \sim p(\mathbf{x},\mathbf{y})} \mathbb{E}_{\tilde{\mathbf{v}}_{x},\mathbf{v}_{y},\mathbf{w},  \tilde{\mathbf{z}}_x, \mathbf{z}_y \sim q_{\boldsymbol{\phi}}}  \Bigg( \Big( \log q_{\boldsymbol{\phi}}(\tilde{\mathbf{v}}_{x}, \tilde{\mathbf{z}}_x \mid \mathbf{x};\phi_x) +  \log q_{\boldsymbol{\phi}}(\mathbf{v}_{y}, \mathbf{z}_y \mid \mathbf{y}; \phi_y) + \log q_{\boldsymbol{\phi}}(\mathbf{w} \mid \mathbf{y}; \phi_f) \Big)  \nonumber  \\
   &   - \Big( \log p_{\boldsymbol{\theta}}(\mathbf{x} \mid \mathbf{w}, \tilde{\mathbf{v}}_{x};\theta_x) 
   +  \log p_{\boldsymbol{\theta}}(\mathbf{y} \mid \mathbf{w}, \mathbf{v}_{y};\theta_y) + \log p(\mathbf{w}) +  \log p(\tilde{\mathbf{v}}_{x} \mid \tilde{\mathbf{z}}_x) + \log p(\mathbf{v}_{y} \mid \mathbf{z}_y) \nonumber \\
    & \hspace{3.5cm} + \underbrace{\log p(\tilde{\mathbf{z}}_x)}_{R_{z_x}} + \underbrace{\log p(\mathbf{z}_y)}_{R_{z_y}} \Big) \Bigg) + const,
\end{align}
where the term $R_{z_x}$ corresponds to the rate of the hyperprior associated with the input image, while the term $R_{z_y}$ corresponds to the rate of the hyperprior associated with the correlated image.
\subsection{DSIN limitations}
\begin{figure}[H]
\centering
\begin{subfigure}[t]{0.25\textwidth}
  \centering
  \includegraphics[width=0.95\linewidth]{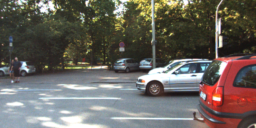}
  \caption{Original Image}
  \label{fig:dsin1}
\end{subfigure}%
\begin{subfigure}[t]{0.25\textwidth}
  \centering
  \includegraphics[width=0.95\linewidth]{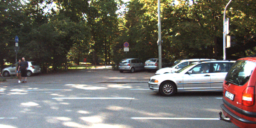}
  \caption{Side Information}
  \label{fig:dsin2}
\end{subfigure}%
\begin{subfigure}[t]{0.25\textwidth}
  \centering
  \includegraphics[width=0.95\linewidth]{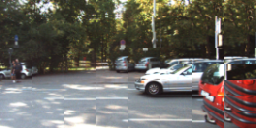}
  \caption{$Y_{syn}$}
  \label{fig:dsin3}
\end{subfigure}%
\begin{subfigure}[t]{0.25\textwidth}
  \centering
  \includegraphics[width=0.95\linewidth]{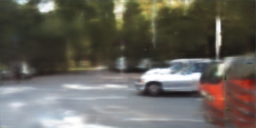}
  \caption{\footnotesize{Reconstructed Image}}
  \label{fig:dsin4}
\end{subfigure}%
\caption{Illustration of the distortion caused by the patch finder in DSIN.}
\label{fig:dsin_images}
\end{figure}

In this section, we illustrate the reason why the DSIN model causes distortions in the image reconstructions. An image $Y_{syn}$ is assembled by the patch finder in the DSIN model by first reconstructing two intermediate images from the input image and the side information from their latent representations, and then finding the most correlated corresponding patches in the two intermediate images. From the offsets of the corresponding patches, the corresponding patches are then taken from the original side information image, and the image $Y_{syn}$ is assembled. In Fig.~\ref{fig:dsin3}, we illustrate that if the intermediate reconstructions of the image and the side information are not good, the patch finder may recognize wrong patches to be the most correlated ones, thus leading to the distorted image $Y_{syn}$ seen in the figure. This leads to an image reconstruction in which patches of the image seem to get blurred or shifted.

\vspace{2cm}

\subsection{Evaluation on full-sized images}
\begin{figure}[H]
\centering
\begin{subfigure}[t]{1\textwidth}
	\centering
	\includegraphics[scale=0.3]{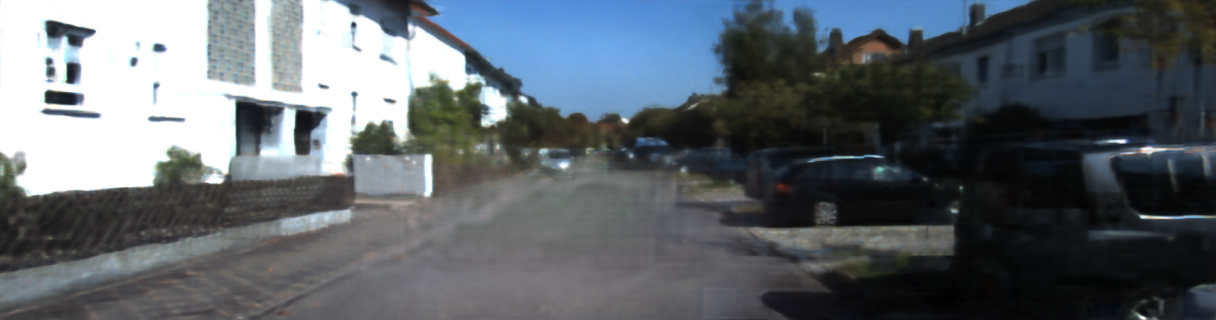}  
	\caption{DSIN, bpp = 0.0435}
	\label{fig:bigger_img_example10}
\end{subfigure}
\vfill
\begin{subfigure}[t]{1\textwidth}
	\centering
	\includegraphics[scale=0.3]{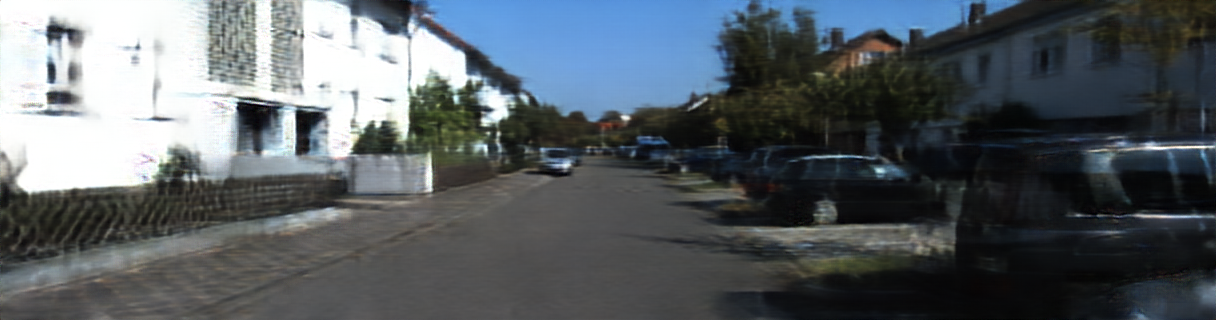}
	\caption{Ours, bpp = 0.0437}
	\label{fig:bigger_img_example11}
\end{subfigure}
\caption{Reconstruction comparison between DSIN (top) and ours (bottom) when evaluated on full-sized images from KITTI Stereo dataset. Compare the texture details captures on the pavement and on the white building.}
\label{fig:bigger_img_example1}
\end{figure}

\begin{figure}[H]
\centering
\begin{subfigure}[]{1\textwidth}
	\centering
	\includegraphics[scale=0.3]{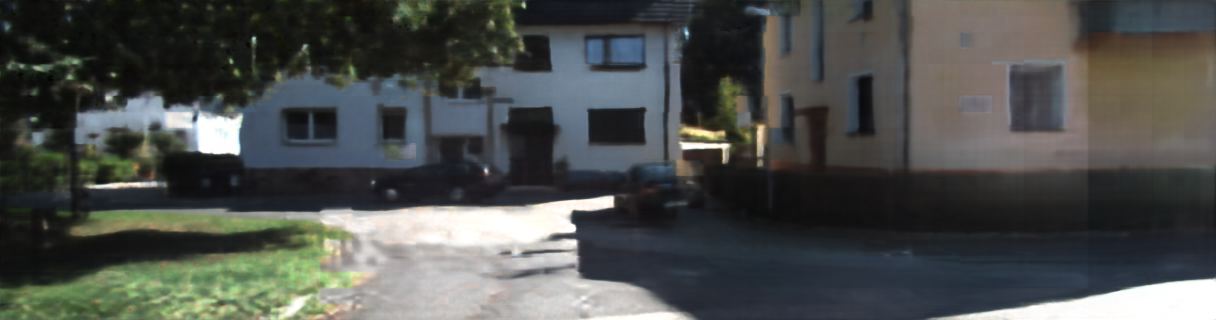} 
	\caption{DSIN, bpp = 0.0449}
	\label{fig:bigger_img_example20}
\end{subfigure}
\vfill
\begin{subfigure}[]{1\textwidth}
	\centering
	\includegraphics[scale=0.3]{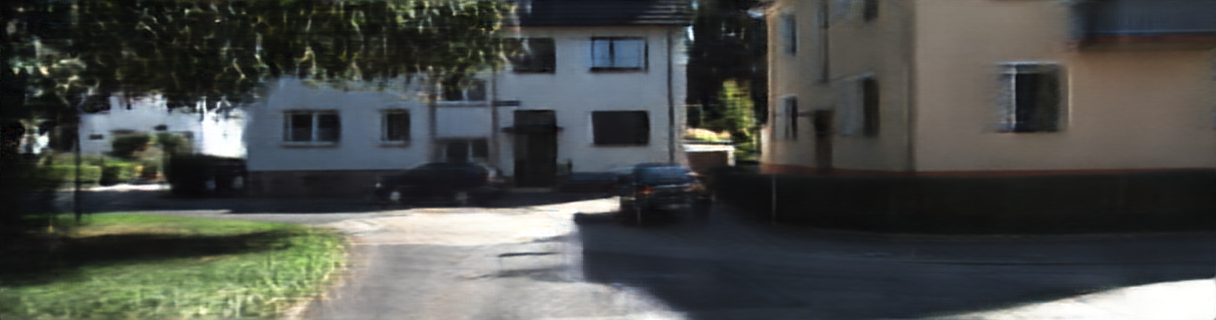}
	\caption{Ours, bpp = 0.0431}
	\label{fig:bigger_img_example21}
\end{subfigure}
\caption{Reconstruction comparison between DSIN (top) and ours (bottom) when evaluated on full-sized images from KITTI Stereo dataset. Compare the edges of the windows, the fine and texture details of the grass and the tree.}
\label{fig:bigger_img_example2}
\end{figure}

In Fig.~\ref{fig:bigger_img_example1} and ~\ref{fig:bigger_img_example2}, we evaluate our trained models on $375 \times 1242$ images from KITTI Stereo dataset. It is interesting to note that although we trained our model and the DSIN model on smaller images of size $128 \times 256$, they perform well on larger images too. We see a significant improvement in the reconstructed images by our proposed model over DSIN, especially in capturing the details of objects which are far away. This is because farther objects do not shift as much as the closer objects in stereo images, and therefore are captured much better as the common information between the stereo images.

\subsection{Additional visual examples -- KITTI Stereo dataset}

\begin{figure}[H]
\captionsetup[subfigure]{labelformat=empty}
\centering
\begin{subfigure}[t]{0.25\textwidth}
	\centering
	\includegraphics[width=0.95\linewidth]{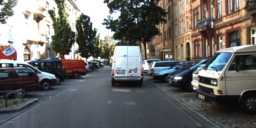}
	\caption{Original Image}
	\label{fig:app0_0}
\end{subfigure}%
\begin{subfigure}[t]{0.25\textwidth}
	\centering
	\includegraphics[width=0.95\linewidth]{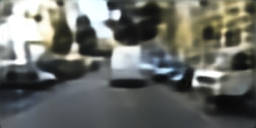}
	\caption{\footnotesize{Ballé2018, bpp = 0.0762} 
	}
	\label{fig:app1_0}
\end{subfigure}%
\begin{subfigure}[t]{0.25\textwidth}
	\centering
	\includegraphics[width=0.95\linewidth]{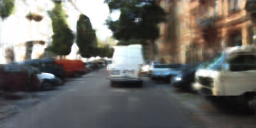}
	\caption{DSIN, bpp = 0.0625
	}
	\label{fig:app3_0}
\end{subfigure}%
\begin{subfigure}[t]{0.25\textwidth}
	\centering
	\includegraphics[width=0.95\linewidth]{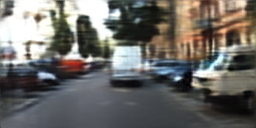}
	\caption{Ours, bpp = 0.0562
	}
	\label{fig:app2_0}
\end{subfigure}%

\begin{subfigure}[t]{0.25\textwidth}
	\centering
	\includegraphics[width=0.95\linewidth]{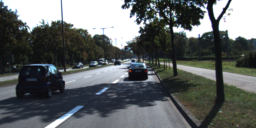}
	\caption{Original Image}
	\label{fig:app0_1}
\end{subfigure}%
\begin{subfigure}[t]{0.25\textwidth}
	\centering
	\includegraphics[width=0.95\linewidth]{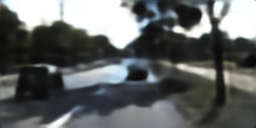}
	\caption{\footnotesize{Ballé2018, bpp = 0.0781}
	}
	\label{fig:app1_1}
\end{subfigure}%
\begin{subfigure}[t]{0.25\textwidth}
	\centering
	\includegraphics[width=0.95\linewidth]{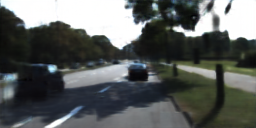}
	\caption{DSIN, bpp = 0.0573
	}
	\label{fig:app3_1}
\end{subfigure}%
\begin{subfigure}[t]{0.25\textwidth}
	\centering
	\includegraphics[width=0.95\linewidth]{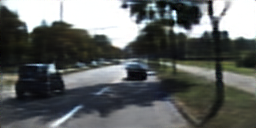}
	\caption{Ours, bpp = 0.0532
	}
	\label{fig:app2_1}
\end{subfigure}%

\begin{subfigure}[t]{0.25\textwidth}
	\centering
	\includegraphics[width=0.95\linewidth]{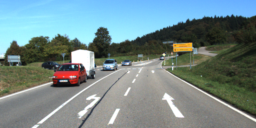}
	\caption{Original Image}
	\label{fig:app0_3}
\end{subfigure}%
\begin{subfigure}[t]{0.25\textwidth}
	\centering
	\includegraphics[width=0.95\linewidth]{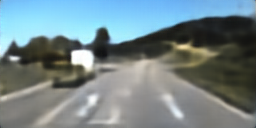}
	\caption{\footnotesize{Ballé2018, bpp = 0.0781} 
	}
	\label{fig:app1_3}
\end{subfigure}%
\begin{subfigure}[t]{0.25\textwidth}
	\centering
	\includegraphics[width=0.95\linewidth]{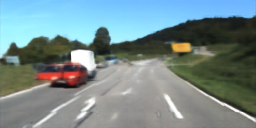}
	\caption{DSIN, bpp = 0.0552
	}
	\label{fig:app3_3}
\end{subfigure}%
\begin{subfigure}[t]{0.25\textwidth}
	\centering
	\includegraphics[width=0.95\linewidth]{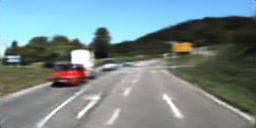}
	\caption{Ours, bpp = 0.0430
	}
	\label{fig:app2_3}
\end{subfigure}%

\begin{subfigure}[t]{0.25\textwidth}
	\centering
	\includegraphics[width=0.95\linewidth]{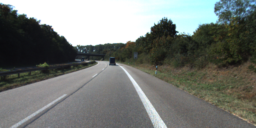}
	\caption{Original Image}
	\label{fig:app0_4}
\end{subfigure}%
\begin{subfigure}[t]{0.25\textwidth}
	\centering
	\includegraphics[width=0.95\linewidth]{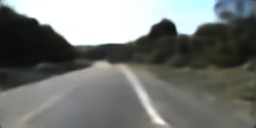}
	\caption{\footnotesize{Ballé2018, bpp = 0.0603}
	}
	\label{fig:app1_4}
\end{subfigure}%
\begin{subfigure}[t]{0.25\textwidth}
	\centering
	\includegraphics[width=0.95\linewidth]{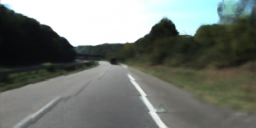}
	\caption{DSIN, bpp = 0.0488
	}
	\label{fig:app3_4}
\end{subfigure}%
\begin{subfigure}[t]{0.25\textwidth}
	\centering
	\includegraphics[width=0.95\linewidth]{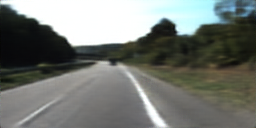}
	\caption{Ours, bpp = 0.0384
	}
	\label{fig:app2_4}
\end{subfigure}%

\begin{subfigure}[t]{0.25\textwidth}
	\centering
	\includegraphics[width=0.95\linewidth]{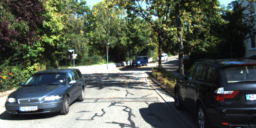}
	\caption{Original Image}
	\label{fig:app0_5}
\end{subfigure}%
\begin{subfigure}[t]{0.25\textwidth}
	\centering
	\includegraphics[width=0.95\linewidth]{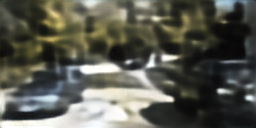}
	\caption{\footnotesize{Ballé2018, bpp = 0.0844}
	}
	\label{fig:app1_5}
\end{subfigure}%
\begin{subfigure}[t]{0.25\textwidth}
	\centering
	\includegraphics[width=0.95\linewidth]{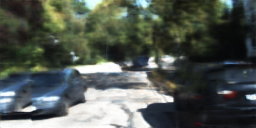}
	\caption{DSIN, bpp = 0.0615
	}
	\label{fig:app3_5}
\end{subfigure}%
\begin{subfigure}[t]{0.25\textwidth}
	\centering
	\includegraphics[width=0.95\linewidth]{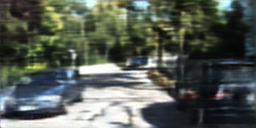}
	\caption{Ours, bpp = 0.0612
	}
	\label{fig:app2_5}
\end{subfigure}%

\begin{subfigure}[t]{0.25\textwidth}
	\centering
	\includegraphics[width=0.95\linewidth]{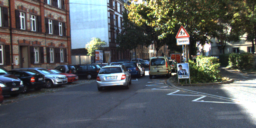}
	\caption{Original Image}
	\label{fig:app0_9}
\end{subfigure}%
\begin{subfigure}[t]{0.25\textwidth}
	\centering
	\includegraphics[width=0.95\linewidth]{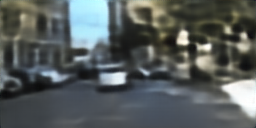}
	\caption{\footnotesize{Ballé2018, bpp = 0.0750}
	}
	\label{fig:app1_9}
\end{subfigure}%
\begin{subfigure}[t]{0.25\textwidth}
	\centering
	\includegraphics[width=0.95\linewidth]{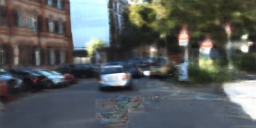}
	\caption{DSIN, bpp = 0.0623
	}
	\label{fig:app3_9}
\end{subfigure}%
\begin{subfigure}[t]{0.25\textwidth}
	\centering
	\includegraphics[width=0.95\linewidth]{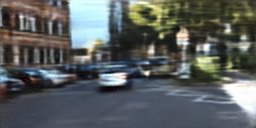}
	\caption{Ours, bpp = 0.0543
	}
	\label{fig:app2_9}
\end{subfigure}%
\caption{Additional visual comparison of different models from KITTI Stereo dataset in a low bpp range. The models are trained using MS-SSIM distortion function. Notice that in some examples, such as rows 3, 4, and 6, the DSIN model adds some fracture-like effects to the reconstructed image.}
\label{fig:vis_exm}
\end{figure}

\begin{figure}[H]
\captionsetup[subfigure]{labelformat=empty}
\centering
\begin{subfigure}[t]{0.25\textwidth}
	\centering
	\includegraphics[width=0.95\linewidth]{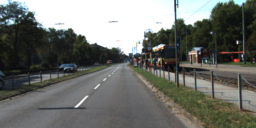}
	\caption{Original Image}
	\label{fig:app2_0_0}
\end{subfigure}%
\begin{subfigure}[t]{0.25\textwidth}
	\centering
	\includegraphics[width=0.95\linewidth]{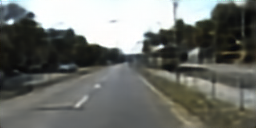}
	\caption{\footnotesize{Ballé2018, bpp = 0.1429}}
	\label{fig:app2_1_0}
\end{subfigure}%
\begin{subfigure}[t]{0.25\textwidth}
	\centering
	\includegraphics[width=0.95\linewidth]{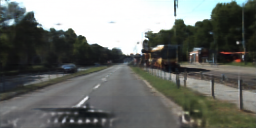}
	\caption{DSIN, bpp = 0.1724}
	\label{fig:app2_3_0}
\end{subfigure}%
\begin{subfigure}[t]{0.25\textwidth}
	\centering
	\includegraphics[width=0.95\linewidth]{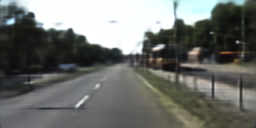}
	\caption{Ours, bpp = 0.1431}
	\label{fig:app2_2_0}
\end{subfigure}%

\begin{subfigure}[t]{0.25\textwidth}
	\centering
	\includegraphics[width=0.95\linewidth]{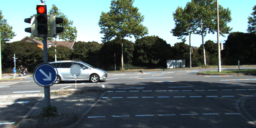}
	\caption{Original Image}
	\label{fig:app2_0_1}
\end{subfigure}%
\begin{subfigure}[t]{0.25\textwidth}
	\centering
	\includegraphics[width=0.95\linewidth]{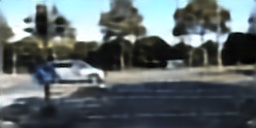}
	\caption{\footnotesize{Ballé2018, bpp = 0.1627}}
	\label{fig:app2_1_1}
\end{subfigure}%
\begin{subfigure}[t]{0.25\textwidth}
	\centering
	\includegraphics[width=0.95\linewidth]{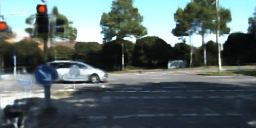}
	\caption{DSIN, bpp = 0.1516}
	\label{fig:app2_3_1}
\end{subfigure}%
\begin{subfigure}[t]{0.25\textwidth}
	\centering
	\includegraphics[width=0.95\linewidth]{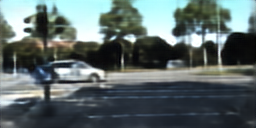}
	\caption{Ours, bpp = 0.1531}
	\label{fig:app2_2_1}
\end{subfigure}%

\begin{subfigure}[t]{0.25\textwidth}
	\centering
	\includegraphics[width=0.95\linewidth]{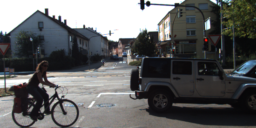}
	\caption{Original Image}
	\label{fig:app2_0_2}
\end{subfigure}%
\begin{subfigure}[t]{0.25\textwidth}
	\centering
	\includegraphics[width=0.95\linewidth]{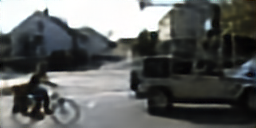}
	\caption{\footnotesize{Ballé2018, bpp = 0.1686}}
	\label{fig:app2_1_2}
\end{subfigure}%
\begin{subfigure}[t]{0.25\textwidth}
	\centering
	\includegraphics[width=0.95\linewidth]{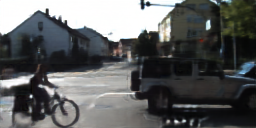}
	\caption{DSIN, bpp = 0.1591}
	\label{fig:app2_3_2}
\end{subfigure}%
\begin{subfigure}[t]{0.25\textwidth}
	\centering
	\includegraphics[width=0.95\linewidth]{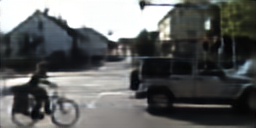}
	\caption{Ours, bpp = 0.1562}
	\label{fig:app2_2_2}
\end{subfigure}%

\begin{subfigure}[t]{0.25\textwidth}
	\centering
	\includegraphics[width=0.95\linewidth]{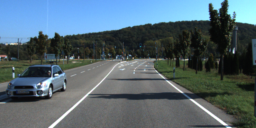}
	\caption{Original Image}
	\label{fig:app2_0_3}
\end{subfigure}%
\begin{subfigure}[t]{0.25\textwidth}
	\centering
	\includegraphics[width=0.95\linewidth]{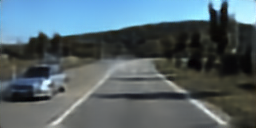}
	\caption{\footnotesize{Ballé2018, bpp = 0.1412}}
	\label{fig:app2_1_3}
\end{subfigure}%
\begin{subfigure}[t]{0.25\textwidth}
	\centering
	\includegraphics[width=0.95\linewidth]{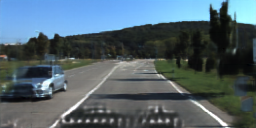}
	\caption{DSIN, bpp = 0.1758}
	\label{fig:app2_3_3}
\end{subfigure}%
\begin{subfigure}[t]{0.25\textwidth}
	\centering
	\includegraphics[width=0.95\linewidth]{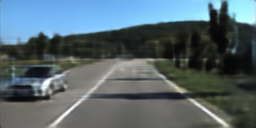}
	\caption{Ours, bpp = 0.1362}
	\label{fig:app2_2_3}
\end{subfigure}%

\begin{subfigure}[t]{0.25\textwidth}
	\centering
	\includegraphics[width=0.95\linewidth]{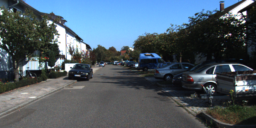}
	\caption{Original Image}
	\label{fig:app2_0_5}
\end{subfigure}%
\begin{subfigure}[t]{0.25\textwidth}
	\centering
	\includegraphics[width=0.95\linewidth]{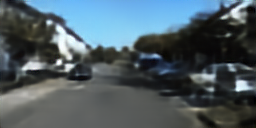}
	\caption{\footnotesize{Ballé2018, bpp = 0.1480}}
	\label{fig:app2_1_5}
\end{subfigure}%
\begin{subfigure}[t]{0.25\textwidth}
	\centering
	\includegraphics[width=0.95\linewidth]{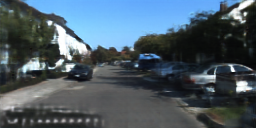}
	\caption{DSIN, bpp = 0.1670}
	\label{fig:app2_3_5}
\end{subfigure}%
\begin{subfigure}[t]{0.25\textwidth}
	\centering
	\includegraphics[width=0.95\linewidth]{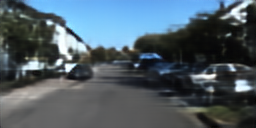}
	\caption{Ours, bpp = 0.1467}
	\label{fig:app2_2_5}
\end{subfigure}%

\begin{subfigure}[t]{0.25\textwidth}
	\centering
	\includegraphics[width=0.95\linewidth]{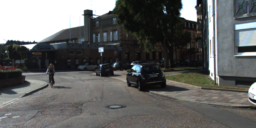}
	\caption{Original Image}
	\label{fig:app2_0_6}
\end{subfigure}%
\begin{subfigure}[t]{0.25\textwidth}
	\centering
	\includegraphics[width=0.95\linewidth]{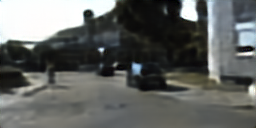}
	\caption{\footnotesize{Ballé2018, bpp = 0.1448}}
	\label{fig:app2_1_6}
\end{subfigure}%
\begin{subfigure}[t]{0.25\textwidth}
	\centering
	\includegraphics[width=0.95\linewidth]{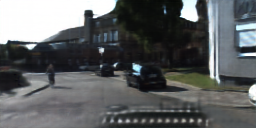}
	\caption{DSIN, bpp = 0.1726}
	\label{fig:app2_3_6}
\end{subfigure}%
\begin{subfigure}[t]{0.25\textwidth}
	\centering
	\includegraphics[width=0.95\linewidth]{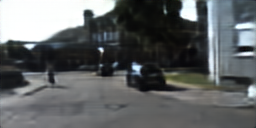}
	\caption{Ours, bpp = 0.1408}
	\label{fig:app2_2_6}
\end{subfigure}%
\caption{Additional visual comparison of different models from KITTI Stereo dataset in a higher bpp range. The models are trained using MS-SSIM distortion function. Notice that while DSIN provides good quality reconstructions, it also introduces foreign artifacts into the images.}
\label{fig:vis_exm2}
\end{figure}

\begin{figure}[H]
\captionsetup[subfigure]{labelformat=empty}
\centering
\captionsetup{justification=centering}
\begin{subfigure}[t]{0.25\textwidth}
	\centering
	\includegraphics[width=0.95\linewidth]{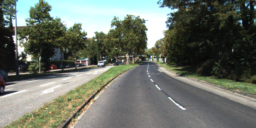}
	\caption{Original Image}
	\label{fig:app3_0_0}
\end{subfigure}%
\begin{subfigure}[t]{0.25\textwidth}
	\centering
	\includegraphics[width=0.95\linewidth]{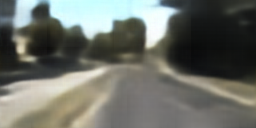}
	\caption{\footnotesize{Ballé2018, bpp = 0.0369}}
	\label{fig:app3_1_0}
\end{subfigure}%
\begin{subfigure}[t]{0.25\textwidth}
	\centering
	\includegraphics[width=0.95\linewidth]{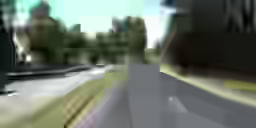}
	\caption{BPG, bpp = 0.0725}
	\label{fig:app3_3_0}
\end{subfigure}%
\begin{subfigure}[t]{0.25\textwidth}
	\centering
	\includegraphics[width=0.95\linewidth]{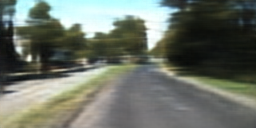}
	\caption{Ours, bpp = 0.0210}
	\label{fig:app3_2_0}
\end{subfigure}%

\begin{subfigure}[t]{0.25\textwidth}
	\centering
	\includegraphics[width=0.95\linewidth]{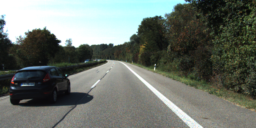}
	\caption{Original Image}
	\label{fig:app3_0_2}
\end{subfigure}%
\begin{subfigure}[t]{0.25\textwidth}
	\centering
	\includegraphics[width=0.95\linewidth]{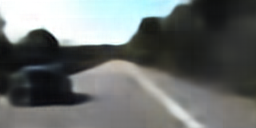}
	\caption{\footnotesize{Ballé2018, bpp = 0.0334}}
	\label{fig:app3_1_2}
\end{subfigure}%
\begin{subfigure}[t]{0.25\textwidth}
	\centering
	\includegraphics[width=0.95\linewidth]{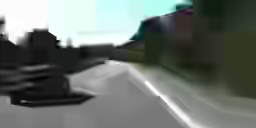}
	\caption{BPG, bpp = 0.0518}
	\label{fig:app3_3_2}
\end{subfigure}%
\begin{subfigure}[t]{0.25\textwidth}
	\centering
	\includegraphics[width=0.95\linewidth]{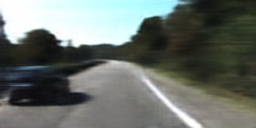}
	\caption{Ours, bpp = 0.0190}
	\label{fig:app3_2_2}
\end{subfigure}%

\begin{subfigure}[t]{0.25\textwidth}
	\centering
	\includegraphics[width=0.95\linewidth]{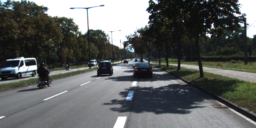}
	\caption{Original Image}
	\label{fig:app3_0_3}
\end{subfigure}%
\begin{subfigure}[t]{0.25\textwidth}
	\centering
	\includegraphics[width=0.95\linewidth]{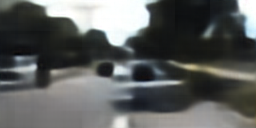}
	\caption{\footnotesize{Ballé2018, bpp = 0.0399}}
	\label{fig:app3_1_3}
\end{subfigure}%
\begin{subfigure}[t]{0.25\textwidth}
	\centering
	\includegraphics[width=0.95\linewidth]{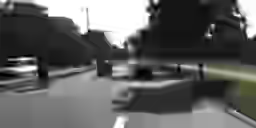}
	\caption{BPG, bpp = 0.0789}
	\label{fig:app3_3_3}
\end{subfigure}%
\begin{subfigure}[t]{0.25\textwidth}
	\centering
	\includegraphics[width=0.95\linewidth]{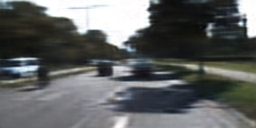}
	\caption{Ours, bpp = 0.0217}
	\label{fig:app3_2_3}
\end{subfigure}%

\begin{subfigure}[t]{0.25\textwidth}
	\centering
	\includegraphics[width=0.95\linewidth]{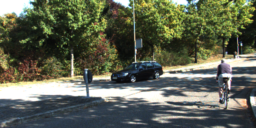}
	\caption{Original Image}
	\label{fig:app3_0_4}
\end{subfigure}%
\begin{subfigure}[t]{0.25\textwidth}
	\centering
	\includegraphics[width=0.95\linewidth]{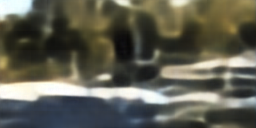}
	\caption{\footnotesize{Ballé2018, bpp = 0.0441}}
	\label{fig:app3_1_4}
\end{subfigure}%
\begin{subfigure}[t]{0.25\textwidth}
	\centering
	\includegraphics[width=0.95\linewidth]{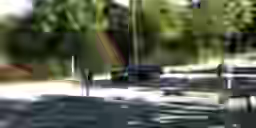}
	\caption{BPG, bpp = 0.1157}
	\label{fig:app3_3_4}
\end{subfigure}%
\begin{subfigure}[t]{0.25\textwidth}
	\centering
	\includegraphics[width=0.95\linewidth]{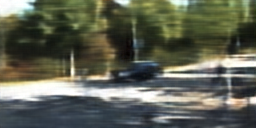}
	\caption{Ours, bpp = 0.0259}
	\label{fig:app3_2_4}
\end{subfigure}%

\begin{subfigure}[t]{0.25\textwidth}
	\centering
	\includegraphics[width=0.95\linewidth]{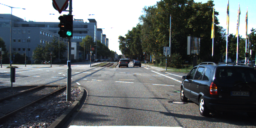}
	\caption{Original Image}
	\label{fig:app3_0_5}
\end{subfigure}%
\begin{subfigure}[t]{0.25\textwidth}
	\centering
	\includegraphics[width=0.95\linewidth]{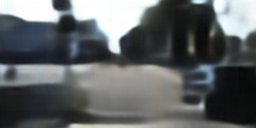}
	\caption{\footnotesize{Ballé2018, bpp = 0.0438}}
	\label{fig:app3_1_5}
\end{subfigure}%
\begin{subfigure}[t]{0.25\textwidth}
	\centering
	\includegraphics[width=0.95\linewidth]{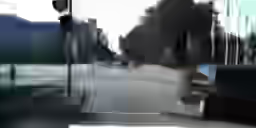}
	\caption{BPG, bpp = 0.0869}
	\label{fig:app3_3_5}
\end{subfigure}%
\begin{subfigure}[t]{0.25\textwidth}
	\centering
	\includegraphics[width=0.95\linewidth]{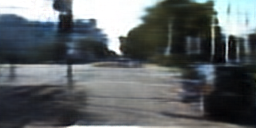}
	\caption{Ours, bpp = 0.0277}
	\label{fig:app3_2_5}
\end{subfigure}%

\begin{subfigure}[t]{0.25\textwidth}
	\centering
	\includegraphics[width=0.95\linewidth]{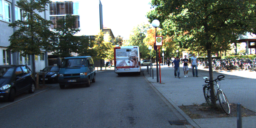}
	\caption{Original Image}
	\label{fig:app3_0_6}
\end{subfigure}%
\begin{subfigure}[t]{0.25\textwidth}
	\centering
	\includegraphics[width=0.95\linewidth]{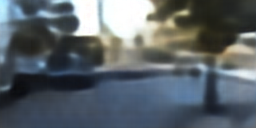}
	\caption{\footnotesize{Ballé2018, bpp = 0.0403}}
	\label{fig:app3_1_6}
\end{subfigure}%
\begin{subfigure}[t]{0.25\textwidth}
	\centering
	\includegraphics[width=0.95\linewidth]{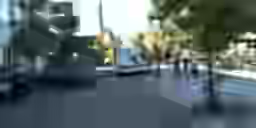}
	\caption{BPG, bpp = 0.0947}
	\label{fig:app3_3_6}
\end{subfigure}%
\begin{subfigure}[t]{0.25\textwidth}
	\centering
	\includegraphics[width=0.95\linewidth]{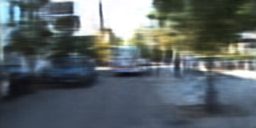}
	\caption{Ours, bpp = 0.0248}
	\label{fig:app3_2_6}
\end{subfigure}%

\caption{Additional visual comparison of different models from KITTI stereo image dataset. The models are trained using the MSE distortion function. For MSE distortion, DSIN did not perform as well as the other models. Thus BPG is used for compression rather than DSIN. Compared to our model, BPG fails to provide details such as texture and edges in the reconstructed image despite having a higher bpp. Similar to DSIN, BPG also introduces some artifacts into the images. Also note that BPG method overall fails to reach the very low bit rates as seen in Fig.~\ref{fig:psnr} and \ref{fig:msssim}.}
\label{fig:vis_exm3}
\end{figure}

\newpage 

\subsection{Additional visual examples -- Cityscape dataset}

\begin{figure}[H]
\captionsetup[subfigure]{labelformat=empty}
\centering
\begin{subfigure}[t]{0.25\textwidth}
	\centering
	\includegraphics[width=0.95\linewidth]{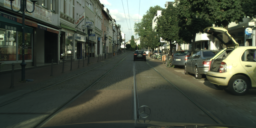}
	\caption{Original Image}
\end{subfigure}%
\begin{subfigure}[t]{0.25\textwidth}
	\centering
	\includegraphics[width=0.95\linewidth]{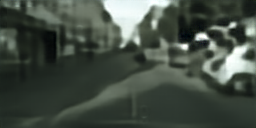}
	\caption{\footnotesize{Ballé2018, bpp = 0.0763}}
\end{subfigure}%
\begin{subfigure}[t]{0.25\textwidth}
	\centering
	\includegraphics[width=0.95\linewidth]{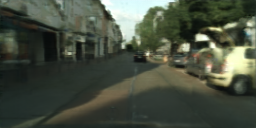}
	\caption{DSIN, bpp = 0.0793}
\end{subfigure}%
\begin{subfigure}[t]{0.25\textwidth}
	\centering
	\includegraphics[width=0.95\linewidth]{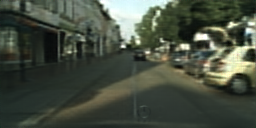}
	\caption{Ours, bpp = 0.0334}
\end{subfigure}%
\\
\begin{subfigure}[t]{0.25\textwidth}
	\centering
	\includegraphics[width=0.95\linewidth]{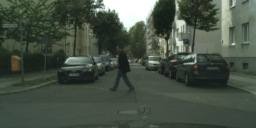}
	\caption{Original Image}
\end{subfigure}%
\begin{subfigure}[t]{0.25\textwidth}
	\centering
	\includegraphics[width=0.95\linewidth]{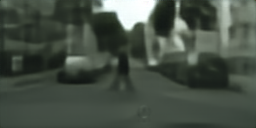}
	\caption{\footnotesize{Ballé2018, bpp = 0.0683}}
\end{subfigure}%
\begin{subfigure}[t]{0.25\textwidth}
	\centering
	\includegraphics[width=0.95\linewidth]{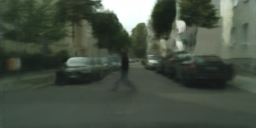}
	\caption{DSIN, bpp = 0.0440}
\end{subfigure}%
\begin{subfigure}[t]{0.25\textwidth}
	\centering
	\includegraphics[width=0.95\linewidth]{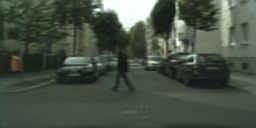}
	\caption{Ours, bpp = 0.0284}
\end{subfigure} 
\begin{subfigure}[t]{0.25\textwidth}
	\centering
	\includegraphics[width=0.95\linewidth]{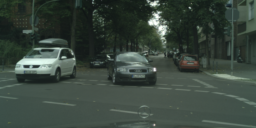}
	\caption{Original Image}
\end{subfigure}%
\begin{subfigure}[t]{0.25\textwidth}
	\centering
	\includegraphics[width=0.95\linewidth]{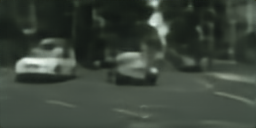}
	\caption{\footnotesize{Ballé2018, bpp = 0.0688}}
\end{subfigure}%
\begin{subfigure}[t]{0.25\textwidth}
	\centering
	\includegraphics[width=0.95\linewidth]{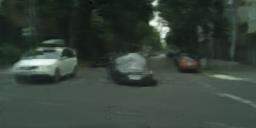}
	\caption{DSIN, bpp = 0.0435}
\end{subfigure}%
\begin{subfigure}[t]{0.25\textwidth}
	\centering
	\includegraphics[width=0.95\linewidth]{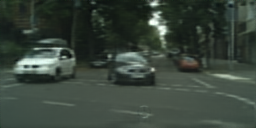}
	\caption{Ours, bpp = 0.0276}
	\end{subfigure}
\begin{subfigure}[t]{0.25\textwidth}
	\centering
	\includegraphics[width=0.95\linewidth]{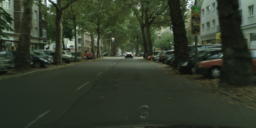}
	\caption{Original Image}
\end{subfigure}%
\begin{subfigure}[t]{0.25\textwidth}
	\centering
	\includegraphics[width=0.95\linewidth]{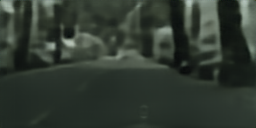}
	\caption{\footnotesize{Ballé2018, bpp = 0.0754}}
\end{subfigure}%
\begin{subfigure}[t]{0.25\textwidth}
	\centering
	\includegraphics[width=0.95\linewidth]{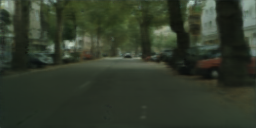}
	\caption{DSIN, bpp = 0.0415}
\end{subfigure}%
\begin{subfigure}[t]{0.25\textwidth}
	\centering
	\includegraphics[width=0.95\linewidth]{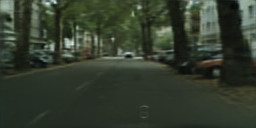}
	\caption{Ours, bpp = 0.0308}
	\end{subfigure}
	\\
\begin{subfigure}[t]{0.25\textwidth}
	\centering
	\includegraphics[width=0.95\linewidth]{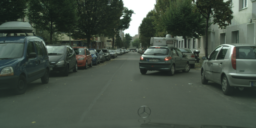}
	\caption{Original Image}
\end{subfigure}%
\begin{subfigure}[t]{0.25\textwidth}
	\centering
	\includegraphics[width=0.95\linewidth]{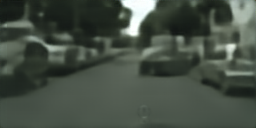}
	\caption{\footnotesize{Ballé2018, bpp = 0.0660}}
\end{subfigure}%
\begin{subfigure}[t]{0.25\textwidth}
	\centering
	\includegraphics[width=0.95\linewidth]{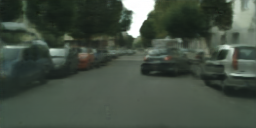}
	\caption{DSIN, bpp = 0.0452}
\end{subfigure}%
\begin{subfigure}[t]{0.25\textwidth}
	\centering
	\includegraphics[width=0.95\linewidth]{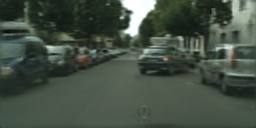}
	\caption{Ours, bpp = 0.0256}
	\end{subfigure}
\begin{subfigure}[t]{0.25\textwidth}
	\centering
	\includegraphics[width=0.95\linewidth]{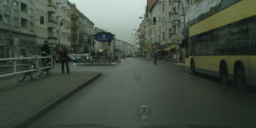}
	\caption{Original Image}
\end{subfigure}%
\begin{subfigure}[t]{0.25\textwidth}
	\centering
	\includegraphics[width=0.95\linewidth]{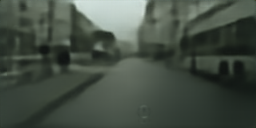}
	\caption{\footnotesize{Ballé2018, bpp = 0.0632}}
\end{subfigure}%
\begin{subfigure}[t]{0.25\textwidth}
	\centering
	\includegraphics[width=0.95\linewidth]{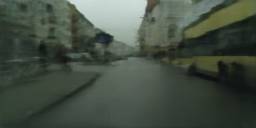}
	\caption{DSIN, bpp = 0.0439}
\end{subfigure}%
\begin{subfigure}[t]{0.25\textwidth}
	\centering
	\includegraphics[width=0.95\linewidth]{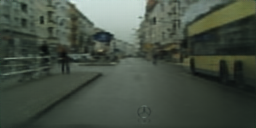}
	\caption{Ours, bpp = 0.02696}
	\end{subfigure}
\caption{Additional visual comparison of different models from Cityscape dataset in a low bpp range. The models are trained using MS-SSIM distortion function.}
\label{fig:vis_exm4}
\end{figure}

\begin{figure}[H]
\captionsetup[subfigure]{labelformat=empty}
\centering
\begin{subfigure}[t]{0.25\textwidth}
	\centering
	\includegraphics[width=0.95\linewidth]{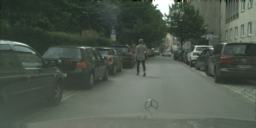}
	\caption{Original Image}
\end{subfigure}%
\begin{subfigure}[t]{0.25\textwidth}
	\centering
	\includegraphics[width=0.95\linewidth]{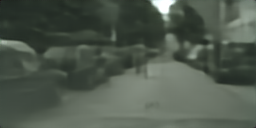}
	\caption{\footnotesize{Ballé2018, bpp = 0.1083}}
\end{subfigure}%
\begin{subfigure}[t]{0.25\textwidth}
	\centering
	\includegraphics[width=0.95\linewidth]{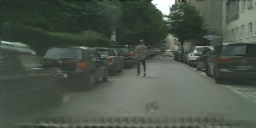}
	\caption{DSIN, bpp = 0.1739}
\end{subfigure}%
\begin{subfigure}[t]{0.25\textwidth}
	\centering
	\includegraphics[width=0.95\linewidth]{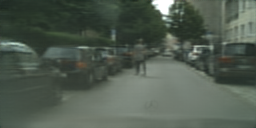}
	\caption{Ours, bpp = 0.0835}
\end{subfigure}

\begin{subfigure}[t]{0.25\textwidth}
	\centering
	\includegraphics[width=0.95\linewidth]{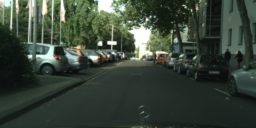}
	\caption{Original Image}
\end{subfigure}%
\begin{subfigure}[t]{0.25\textwidth}
	\centering
	\includegraphics[width=0.95\linewidth]{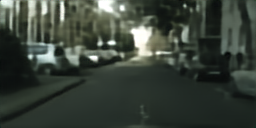}
	\caption{\footnotesize{Ballé2018, bpp = 0.1586}}
\end{subfigure}%
\begin{subfigure}[t]{0.25\textwidth}
	\centering
	\includegraphics[width=0.95\linewidth]{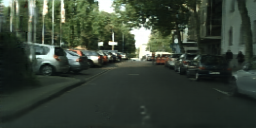}
	\caption{DSIN, bpp = 0.1640}
\end{subfigure}%
\begin{subfigure}[t]{0.25\textwidth}
	\centering
	\includegraphics[width=0.95\linewidth]{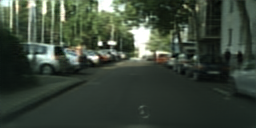}
	\caption{Ours, bpp = 0.1055}
\end{subfigure} \\
\begin{subfigure}[t]{0.25\textwidth}
	\centering
	\includegraphics[width=0.95\linewidth]{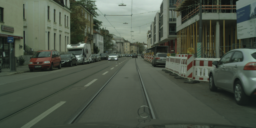}
	\caption{Original Image}
\end{subfigure}%
\begin{subfigure}[t]{0.25\textwidth}
	\centering
	\includegraphics[width=0.95\linewidth]{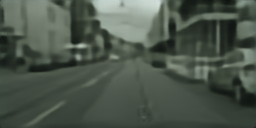}
	\caption{\footnotesize{Ballé2018, bpp = 0.1116}}
\end{subfigure}%
\begin{subfigure}[t]{0.25\textwidth}
	\centering
	\includegraphics[width=0.95\linewidth]{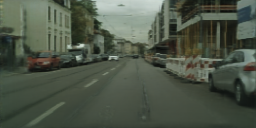}
	\caption{DSIN, bpp = 0.0933}
\end{subfigure}%
\begin{subfigure}[t]{0.25\textwidth}
	\centering
	\includegraphics[width=0.95\linewidth]{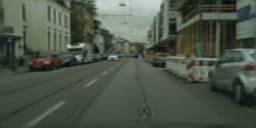}
	\caption{Ours, bpp = 0.0548}
\end{subfigure} 
\\
\begin{subfigure}[t]{0.25\textwidth}
	\centering
	\includegraphics[width=0.95\linewidth]{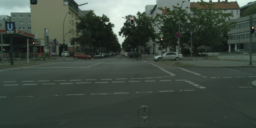}
	\caption{Original Image}
\end{subfigure}%
\begin{subfigure}[t]{0.25\textwidth}
	\centering
	\includegraphics[width=0.95\linewidth]{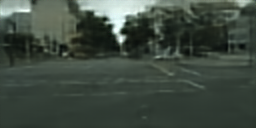}
	\caption{\footnotesize{Ballé2018, bpp = 0.1657}}
\end{subfigure}%
\begin{subfigure}[t]{0.25\textwidth}
	\centering
	\includegraphics[width=0.95\linewidth]{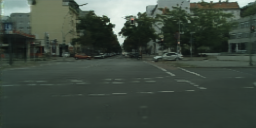}
	\caption{DSIN, bpp = 0.1428}
\end{subfigure}%
\begin{subfigure}[t]{0.25\textwidth}
	\centering
	\includegraphics[width=0.95\linewidth]{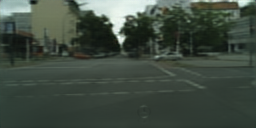}
	\caption{Ours, bpp = 0.0789}
\end{subfigure} \\

\begin{subfigure}[t]{0.25\textwidth}
	\centering
	\includegraphics[width=0.95\linewidth]{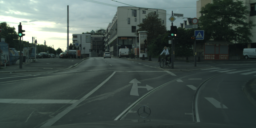}
	\caption{Original Image}
\end{subfigure}%
\begin{subfigure}[t]{0.25\textwidth}
	\centering
	\includegraphics[width=0.95\linewidth]{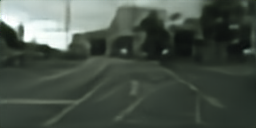}
	\caption{\footnotesize{Ballé2018, bpp = 0.0734}}
\end{subfigure}%
\begin{subfigure}[t]{0.25\textwidth}
	\centering
	\includegraphics[width=0.95\linewidth]{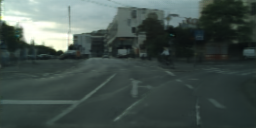}
	\caption{DSIN, bpp = 0.0804}
\end{subfigure}%
\begin{subfigure}[t]{0.25\textwidth}
	\centering
	\includegraphics[width=0.95\linewidth]{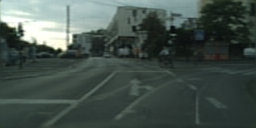}
	\caption{Ours, bpp = 0.0472}
\end{subfigure}\\
\begin{subfigure}[t]{0.25\textwidth}
	\centering
	\includegraphics[width=0.95\linewidth]{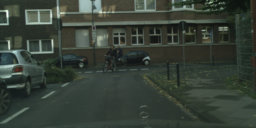}
	\caption{Original Image}
\end{subfigure}%
\begin{subfigure}[t]{0.25\textwidth}
	\centering
	\includegraphics[width=0.95\linewidth]{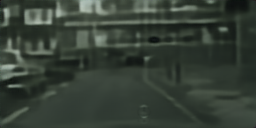}
	\caption{\footnotesize{Ballé2018, bpp = 0.1269}}
\end{subfigure}%
\begin{subfigure}[t]{0.25\textwidth}
	\centering
	\includegraphics[width=0.95\linewidth]{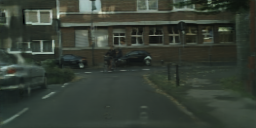}
	\caption{DSIN, bpp = 0.0967}
\end{subfigure}%
\begin{subfigure}[t]{0.25\textwidth}
	\centering
	\includegraphics[width=0.95\linewidth]{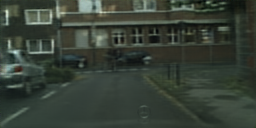}
	\caption{Ours, bpp = 0.0525}
\end{subfigure}%

\caption{Additional visual comparison of different models from Cityscape dataset in a higher bpp range. Notice that while DSIN yields satisfactory quality reconstructions, it also introduces foreign artifacts, and sometimes structural distortions, like breaks in straight lines, into the images.}
\label{fig:vis_exm5}
\end{figure}

\end{document}